\PassOptionsToPackage{dvipsnames}{xcolor}
\documentclass{SciPost}

\hypersetup{
    colorlinks,
    linkcolor={red!50!black},
    citecolor={blue!50!black},
    urlcolor={blue!80!black}
}

\usepackage[bitstream-charter]{mathdesign}
\usepackage{float}
\usepackage[dvipsnames]{xcolor}

\DeclareSymbolFont{usualmathcal}{OMS}{cmsy}{m}{n}
\DeclareSymbolFontAlphabet{\mathcal}{usualmathcal}

\fancypagestyle{SPstyle}{
\fancyhf{}
\lhead{\colorbox{scipostblue}{\bf \color{white} ~SciPost Physics }}
\rhead{{\bf \color{scipostdeepblue} ~Submission }}

\fancyfoot[C]{\textbf{\thepage}}
}

\newcommand{\ket}[1]{\ensuremath{\left| #1 \right>}}

\newcommand{\beg}{\begin{equation}}
\newcommand{\en}{\end{equation}}

\newcommand{\eref}[1]{Eq.~(\ref{#1})}
\newcommand{\re}[1]{(\ref{#1})}

\newcommand{\esref}[1]{Eqs.~(\ref{#1})}

\newcommand{\be}{\begin{equation}}
\newcommand{\ee}{\end{equation}}
\newcommand{\bea}{\begin{eqnarray}}
\newcommand{\eea}{\end{eqnarray}}

\begin{document}

\pagestyle{SPstyle}

\begin{center}{\Large \textbf{\color{scipostdeepblue}{
Exact many-body wavefunction of the Kondo model with time-dependent interaction strength \\
}}}\end{center}

 \begin{center}\textbf{
Parameshwar R. Pasnoori\textsuperscript{1$\star$} and
 Emil A. Yuzbashyan\textsuperscript{2$\dagger$}
}\end{center}

\begin{center}
{\bf 1} Department of Physics, University of Maryland, College Park, MD 20742, USA

{\bf 2}  Department of Physics and Astronomy, Center for Materials Theory, Rutgers University, Piscataway, NJ 08854, USA
\\[\baselineskip]
$\star$ \href{mailto:pparmesh@umd.edu}{\small pparmesh@umd.edu}\,,\quad
$\dagger$ \href{mailto:eyuzbash@rutgers.edu}{\small eyuzbash@rutgers.edu}
\end{center}

\section*{\color{scipostdeepblue}{Abstract}}
\textbf{\boldmath{%
Quantum integrability has been applied to a large variety of low dimensional Hamiltonians in  Quantum Field Theory, Condensed Matter Physics, and Statistical Mechanics to obtain exact expressions for the spectrum and thermodynamics of these systems. In most of these studies the coupling constants are constant in time.  Here we present an exact solution of the nonstationary Schr\"odinger equation for the Kondo Hamiltonian with a time-dependent spin-exchange coupling $J(t)$ of the form $\lambda t + p(t) \pm \sqrt{(\lambda t + p(t))^2 + 16/3}$, where $p(t)$ is an arbitrary periodic function, under periodic boundary conditions. Unlike previously studied time-dependent integrable models, which are rooted in the classical Yang--Baxter structure and associated Knizhnik--Zamolodchikov equations, our approach is based on the quantum Knizhnik--Zamolodchikov   framework and the quantum Yang--Baxter   algebra. Our results broaden the domain of time-dependent integrability to a genuinely quantum class of models and provide new tools for exploring coherent nonequilibrium dynamics in strongly correlated systems.}
}

\vspace{\baselineskip}



\vspace{10pt}
\noindent\rule{\textwidth}{1pt}
\tableofcontents
\noindent\rule{\textwidth}{1pt}
\vspace{10pt}


\section{Introduction}

 In the past, theoretical studies of many-body quantum systems typically began with a specific material or device whose microscopic properties dictated the Hamiltonian. Advances in cold atoms, trapped ions, superconducting qubits, and other programmable platforms have reversed this logic~\cite{GeorgescuRMP2014,GrossBlochScience2017,BlattRoosNatPhys2012,HouckTureciKochNatPhys2012,MonroeRMP2021,MiScience2024,GongScience2021,SatzingerScience2021,CongPRX2022}. We can now engineer Hamiltonians with almost arbitrary interactions and precise time dependence, making the \textsl{model or phenomenon} the starting point and its experimental realization the next step. Integrable models are especially appealing in this context, as they display robust and highly non-generic dynamics---such as the absence of conventional thermalization and the persistence of coherent structures---that can be realized and explored with unprecedented control~\cite{KinoshitaNature2006,RigolPRL2007,FosterPRL2010,PolkovnikovRMP2011,FosterPRB2011,MatsunagaShimanoPRL2012,BeckPRL2013,CauxEsslerPRL2013,MatsunagaScience2014,FosterPRL2014,WoutersPRL2014,YuzbashyanPRA2015,LangenScience2015,IlievskiPRL2015,EsslerFagotti2016,VidmarRigol2016,CastroAlvaredoPRX2016,YuzbashyanAOP2016,MatsunagaPRB2017,ScaramazzaPRB2019,ShankarPRXQ2022,MorvanNature2022,YoungNature2024,KimPolkovnikovPRB2024}.

At the most elementary end of the spectrum lie time-dependent models that  either involve a single degree of freedom or are effectively noninteracting. The simplest well-known examples are the $2\times 2$ Landau--Zener problem~\cite{Landau,zener,majorana,Stueckelberg} and the driven harmonic oscillator~\cite{FeynmanHibbs,MukhanovWinitzki,PerelomovPopov}. Building on the $2\times 2$ case, more involved yet still tractable multilevel Landau--Zener systems were introduced and solved by elementary means, including the Demkov--Osherov and 
bow-tie models~\cite{DemkovOsherov,OstrovskyNakamura,DemkovOstrovsky2000,DemkovOstrovsky2001,ChernyakSinitsynSun2018,ChernyakSinitsynSun,Barik2025}, as well as the  Ising chain in a time-dependent transverse field~\cite{Polkovnikov2005,ZurekDornerZoller2005,Dziarmaga2005}. More recently, similar techniques have been applied to Hamiltonians that are linear combinations of Lie-algebra generators with time-dependent coefficients~\cite{Galitski2011,RingelGritsev2013,DavidsonPolkovnikov2015,PatraYuzbashyan2015,GritsevPolkovnikov,MalikisCheianov2025}. On the classical side, integrable nonlinear equations---such as the nonlinear Schr\"odinger and Korteweg--de Vries equations---with time-dependent parameters have been analyzed~\cite{Hirota1979,Nirmala1986,Chan1989,GangulyDas2015}, but these usually turn out to be related to their time-independent counterparts by a simple change of variables~\cite{Hlavaty1986,Joshi1987,Brugarino1989,PerezGarcia2006}. The tools of classical integrability have likewise been used to address Landau--Zener--type problems~\cite{MalikisCheianov2025,MalikisCheianov2022} and to construct integrable sigma models with time-dependent couplings~\cite{Hoare2020}. All of these cases, however, remain within the scope of problems that can be reduced to single-particle, mean-field, or classical dynamics, far from the complexity of a fully interacting quantum many-body system.

  Explicit time dependence in genuinely  many-body  Hamiltonians presents significant challenges: while exact solutions abound for stationary integrable models, continuous driving typically destroys known integrable structures.
  The emerging field of 
  \textsl{time-dependent integrability}~\cite{PatraYuzbashyan2015,sinitsyn_integrable_2018,Yuzbashyan2018,Zabalo2022,SuzukiMallaSinitsyn2025arxiv} 
 seeks to bridge this gap by identifying nonautonomous Hamiltonians whose  dynamics remain exactly solvable. Thus far, however, the solvable landscape is rather narrow, consisting  primarily of various models that reduce to commuting Gaudin magnets~\cite{GaudinBook,Sklyanin1989,Ortiz2005,Skrypnyk2007b,Skrypnyk2007,FiorettoCauxGritsev2014,Barik2025RG} (central-spin Hamiltonians) and their linear combinations, such as the Bardeen--Cooper--Schrieffer (BCS) model~\cite{sinitsyn_integrable_2018,Yuzbashyan2018,Zabalo2022,LiChernyakSinitsyn2018} and generalized bosonic~\cite{SuzukiMallaSinitsyn2025arxiv,Malla2022} and fermionic~\cite{sinitsyn_integrable_2018,Yuzbashyan2018,Sun2019,SinitsynLi2016,SunSinitsyn2016} Tavis--Cummings (inhomogeneous Dicke) models with time-dependent parameters. These relatively simple constructions descend from the \textsl{classical} Yang--Baxter equation~\cite{Torrielli2016,Loebbert2016,Retore2022}  (the  semiclassical limit of the \textsl{quantum} Yang--Baxter equation) and remain close, in spirit, to the classical or mean-field limits. 

 Here we extend time-dependent integrability to a broader and more complex class
of quantum integrable models derived from the full quantum Yang--Baxter equation~\cite{McGuire1964,Yang1967,Zamolodchikov1979,Baxter1982,Jimbo1989}, continuing the program started by one of the authors in Ref.~\cite{Parmesh1}.
As a concrete example, we obtain the exact solution of the nonstationary Schr\"odinger equation for the Kondo Hamiltonian with a time-dependent spin--exchange coupling \(J(t)\).  
The model exemplifies a genuinely quantum many-body impurity system, distinct from noninteracting, mean-field, or classical limits. 
Here we solve the constraints on integrable $J(t)$ identified in
Ref.~\cite{Parmesh1}, obtain the general family of couplings compatible
with the qKZ mapping, and construct the explicit many-body wavefunction
by solving the qKZ equations. The scattering construction and derivation
of the consistency conditions are reviewed below to make the
presentation self-contained.
Our method applies more generally to any integrable one-dimensional model with linear dispersion, enabling exact analysis of a range of driven quantum many-body systems and establishing a systematic framework for treating exact real-time dynamics in such models.  To our knowledge, this is the first explicit many-body solution of a genuinely quantum impurity model with a nontrivial time-dependent interaction. Our solution opens the way to controlled evaluations of experimentally relevant quantities—including impurity polarization, Knight-shift correlations, entanglement entropy, and momentum distributions.

The Kondo model describes a magnetic impurity with spin-$\frac{1}{2}$ interacting with conduction electrons through a spin-exchange coupling $J$. In thermal equilibrium, it exhibits the Kondo effect: the resistivity increases as the temperature is lowered and saturates at a finite value as $T \to 0$~\cite{Anderson,Wilson,Nozier}. The Hamiltonian reads
\beg
H[J] = -i\! \int\limits_{-L/2}^{L/2} \! {\hat \psi}^{\dagger}_{s}(x) \, \partial_x  {\hat \psi}_{s}(x)\, dx 
+  J\,  {\hat \psi}^{\dagger}_{s}(0) \, \vec{\sigma}_{ss'} \, {\hat \psi}_{s'}(0) \cdot \vec{S},
\label{Hamiltonian}
\en
where ${\hat \psi}_{s}(x)$ is the electron field operator with spin index $s = \uparrow, \downarrow$, $\vec{S}$ is the spin operator of the impurity located at $x = 0$, $\vec{\sigma}$ is the vector of Pauli matrices, and $L$ is the system size. Throughout, we set the Fermi velocity $v_F = 1$ and assume summation over repeated spin indices.

The stationary (time-independent) Schr\"odinger equation for this Hamiltonian has been solved using the Bethe Ansatz~\cite{Andrei80,Wiegmann_1981}. In this work, we consider a time-dependent spin-exchange coupling of the form
\beg
J(t) = \lambda t + p(t) \pm \sqrt{[\lambda t + p(t)]^2 + \tfrac{16}{3}},
\label{fn_form178}
\en
where $p(t)$ is an arbitrary periodic function of period $L$ and $\lambda$ is a real parameter.\footnote{The exact functional form of the time-dependent interaction strength depends on the regularization scheme.}   We solve exactly the nonstationary Schr\"odinger equation with periodic boundary conditions for the corresponding Hamiltonian $H(t) = H[J(t)]$, and show that~\eref{fn_form178} specifies the most general functional form of $J(t)$ for which the problem admits an exact solution. In the diabatic limit $\lambda\to\infty$ \footnote{The characteristic driving time
scales as $1/\lambda$, so $\lambda\to\infty$ is the rapid-driving
limit. The inverse-time form $J(t)\approx C/(a+t)$ applies on
the weak-coupling side, with $C\propto1/\lambda$.} with constant $p(t)$, the coupling reduces to   $J(t)\approx C/(a+t),$ a case previously identified as integrable in  Ref.~\cite{Parmesh1}.

Apart from the arbitrary periodic function $p(t)$, the time-dependent Kondo Hamiltonian considered here contains a single free parameter, $\lambda$, which sets the characteristic time scale. In contrast, most previously studied models involve several additional parameters $z_1, \dots, z_n$; for example, in the BCS model the $z_i$ represent fixed single-particle energy levels~\cite{sinitsyn_integrable_2018,Yuzbashyan2018}. In models derived from the quantum Yang--Baxter equation, however, these static external parameters are promoted to dynamical variables---specifically, the light-cone coordinates of the particles---fundamentally altering the structure of the problem.  Sinitsyn \textit{et al.} proposed in 2017~\cite{sinitsyn_integrable_2018} that a Hamiltonian $H(z_1, \dots, z_n; t)$ is integrable if its nonstationary Schr\"odinger equation is  compatible with a system of evolution equations,  
\begin{equation}
\label{system1}
 i \nu\,\frac{\partial \Psi(\vec{z})}{\partial z_i} = H_{i} \,\Psi(\vec{z}), \quad i = 0, 1, \ldots, n,
\end{equation}
where the Hermitian operators $H_i$ serve as auxiliary Hamiltonians generating evolution with respect to each $z_i$.

The connection between this conjecture and the classical Yang--Baxter equation becomes clear when the auxiliary Hamiltonians $H_i$ are taken in the form
\begin{equation}
\label{system2}
 H_i = \sum_{\substack{j=1 \\ j \neq i}}^n r_{ij}(z_i, z_j), \quad i = 1, \ldots, n,
\end{equation}
where $r_{ij}(u_1,u_2)$ acts in the combined Hilbert space of particles $i$ and $j$ and depends on two complex parameters $u_1$ and $u_2$. Cherednik observed in 1989~\cite{Cherednik1989,Etingof1994}\footnote{Cherednik made the standard assumptions that $r_{ij}(u_1,u_2)$ depends only on the difference of its arguments, $r_{ij}(u_1,u_2)=r_{ij}(u_1-u_2)$, and that it is skew-symmetric, $r_{ij}(u_1-u_2)=-r_{ij}(u_2-u_1)$. Relaxing these assumptions does not appear to substantially alter the overall framework~\cite{Skrypnyk2007b,Torrielli2016}.} that the multi-time Schr\"odinger system~\eqref{system1} is compatible if and only if $r_{ij}(u_1,u_2)$ satisfies the \textsl{classical Yang--Baxter equation}
\beg
[r_{ij}(z_i, z_j), r_{ik}(z_i, z_k)] 
+ [r_{ij}(z_i, z_j), r_{jk}(z_j, z_k)] 
+ [r_{kj}(z_k, z_j), r_{ik}(z_i, z_k)] = 0.
\label{cybe}
\en
Solutions of~\eqref{cybe} have been extensively studied and classified into three families: rational, trigonometric (including hyperbolic), and elliptic~\cite{BelavinDrinfeld1982,BelavinDrinfeld1984,Torrielli2016,Retore2022,Skrypnyk2007b,Stolin1991}. The associated $H_i$ correspond to the rational, trigonometric, and elliptic Gaudin magnets, respectively, and the resulting evolution equations~\eqref{system1} are known as generalized Knizhnik--Zamolodchikov (KZ) equations~\cite{Etingof1994}.

Further generalizations of the KZ equations can be obtained by adding boundary terms to \(H_i\) in \eref{system2}\cite{Hikami1995,BabujianKitaev1998,KurakLimaSantos2004,LimaSantosUtiel2006,FiorettoCauxGritsev2014,SedrakyanGalitski2010,Skrypnyk2010}. As a simple example, consider a rational solution of the classical Yang--Baxter equation~\re{cybe},  
\beg
r_{ij}(z_i, z_j) = \frac{\vec{s}_i \cdot \vec{s}_j}{z_i - z_j},
\label{rij_intro}
\en
where \(\vec{s}_k\) are \({su}(2)\) spin operators. With this choice of \(r_{ij}(z_i, z_j)\), the operators~\re{system2} become the rational, isotropic \({su}(2)\) Gaudin magnets discussed earlier, and the corresponding equations~\re{system1} reduce to the original (i.e., pre-generalization)  KZ differential equations describing \(n\)-point correlation functions \(\Psi(\vec z)\) in two-dimensional conformal field theory~\cite{KnizhnikZamolodchikov1984}.  

A boundary term---corresponding to a Zeeman magnetic field acting on the central spin \(\vec{s}_i\)---can be added to \(H_i\) without spoiling the compatibility of Eqs.~\re{system1}~\cite{FiorettoCauxGritsev2014,SedrakyanGalitski2010}:  
\beg
H_i = B\,s_i^z + \sum_{\substack{j=1 \\ j \neq i}}^n  \frac{\vec{s}_i \cdot \vec{s}_j}{z_i - z_j}, \quad i = 1, \ldots, n.
\label{gaudin}
\en
Choosing the Zeeman field to be linearly proportional to time, \(B = \nu t\), yields a system of multi-time Schr\"odinger equations~\re{system1} compatible with the nonstationary Schr\"odinger equation for the BCS Hamiltonian~\cite{Yuzbashyan2018}, where the superconducting coupling is inversely proportional to time, \(g(t) = 1/(\nu t)\). In the limit where the length of the spin \(\vec{s}_i\) tends to infinity---so that it can be replaced by a harmonic oscillator---this \(H_i\) reduces to the inhomogeneous Dicke (generalized fermionic Tavis--Cummings) model with the bosonic energy level varying linearly in time~\cite{Yuzbashyan2018}. Furthermore, replacing the \(su(2)\) spins \(\vec{s}_{j \neq i}\) in \eref{gaudin} with \(su(1,1)\) spins yields the time-dependent bosonic Tavis--Cummings model studied in Refs.~\cite{SuzukiMallaSinitsyn2025arxiv,Malla2022}. Thus,  previously known integrable time-dependent quantum many-body models can be traced back to this single family of rational Gaudin magnets with a boundary (Zeeman-field) term.

While previously studied time-dependent integrable models are rooted in the \textsl{classical} Yang--Baxter structure and the associated KZ equations, our solution of the time-dependent Kondo model is based instead on the \textsl{quantum} Knizhnik--Zamolodchikov (qKZ) framework~\cite{Frenkel,rishetikhin1,smirnovqKZ,smirnovqkz2,JimboqKZ,Babujian_1997,Etingof1998}
 and the corresponding quantum Yang--Baxter algebra. The qKZ equations form a system of finite-difference equations for a vector-valued function (a state in a certain Hilbert space) $\varphi(y_0,\dots,y_j,\dots,y_N)$,
\beg
\varphi(y_0,\dots,y_j+\kappa,\dots,y_N) 
= M_j(y_0,\dots,y_N)\,\varphi(y_0,\dots,y_j,\dots,y_N),
\label{diffeq12_intro}
\en
with step $\kappa$ and transport operators $M_j$ given by ordered products of quantum $R$-matrices,
\beg
M_j(y_0,\dots,y_N)
= R^{j+1\,j}(y_{j+1}-\kappa,y_j)\cdots R^{N\,j}(y_N-\kappa,y_j)\,
R^{0\,j}(y_0,y_j)\cdots R^{j-1\,j}(y_{j-1},y_j).
\label{qKZ_intro}
\en
The $R$-matrices, which act on the state $\varphi$, satisfy the \textsl{quantum} Yang--Baxter equation  
\beg
R^{ij}(z_i,z_j)\, R^{ik}(z_i,z_k)\, R^{jk}(z_j,z_k)
=
R^{jk}(z_j,z_k)\, R^{ik}(z_i,z_k)\, R^{ij}(z_i,z_j).
\label{QYBE_intro}
\en
For the time-dependent Kondo Hamiltonian, $R^{ij}(z_i,z_j)$ is the XXX $R$-matrix:\footnote{$R^{ij}(y) = \frac{i y\, I^{ij} + c\, P^{ij}}{i y + c}$. This representation follows from~\eref{R_intro} by setting $z_i - z_j = i y$ and using the fact that the permutation operator for two spin-$\tfrac{1}{2}$ degrees of freedom $\vec{s}_i$ and $\vec{s}_j$ can be written as $P^{ij} = \tfrac{1}{2} + 2\, \vec{s}_i \cdot \vec{s}_j$.}
\be
R^{ij}(z_i,z_j) = \frac{z_i-z_j + \frac{c}{2}+ 2c\, \vec s_i\cdot \vec s_j}{z_i-z_j + c},
\label{R_intro}
\en
where $\vec s_k$ are spin-$\frac{1}{2}$ operators and $c$ is the crossing parameter.

The \emph{quasi-classical} limit is obtained by setting
\beg
c = \frac{\hbar}{2},\quad \kappa = i \nu\, \hbar,
\en
and taking $\hbar \to 0$. In this limit~\cite{Torrielli2016,Loebbert2016,Retore2022},
\beg
R^{ij}(z_i,z_j) = 1 + \hbar\, r_{ij}(z_i, z_j) + O(\hbar^2),
\en
where $r_{ij}(z_i, z_j)$ is given by~\eref{rij_intro} and corresponds to the rational Gaudin magnets.  Equation~\re{QYBE_intro} then reduces to the classical Yang--Baxter equation~\re{cybe}, and the qKZ system~\re{qKZ_intro} becomes the corresponding differential KZ equations~\cite{Frenkel,JimboqKZ,Etingof1998,Nakayashiki1999}. 

At the same time, the \textsl{physical meaning} of these equations changes. In the Kondo problem, the qKZ equations encode the periodic boundary conditions for the light-cone coordinates $z_i$ of the particles. By contrast, in Gaudin-type models the $z_i$ are external parameters, and the associated KZ equations describe how the many-body wavefunction depends on them. More generally, for an arbitrary $R$-matrix, the quasi-classical limit yields the classical Yang--Baxter equation~\re{cybe} together with the generalized KZ equations. Notably, integrable systems derived from the classical Yang--Baxter equation tend to have a simpler algebraic structure than those arising from the full quantum version, since classical $r$-matrices form a representation of an infinite-dimensional Lie algebra, whereas the quantum case encodes genuinely nonlinear operator relations~\cite{Semenov1985,Sklyanin1989,Torrielli2016,Loebbert2016,Retore2022}.

 The paper is organized as follows. In the next section, we present a general solution of the nonstationary Schr\"odinger equation for the time-dependent Kondo Hamiltonian. We then impose periodic boundary conditions and analyze the resulting quantization conditions. Section~\ref{sec:mapping_qkz} describes the mapping to the qKZ system and determines the time dependence of the spin-exchange coupling $J(t)$ for which the dynamics is integrable. In Section~\ref{wavefunction_sec}, we construct the exact many-body solution of the nonstationary Schr\"odinger equation for the Kondo Hamiltonian with this $J(t)$. We conclude with a summary and a discussion of possible physical realizations, problems that can be addressed using this solution, and generalizations to other integrable one-dimensional models (including Gross--Neveu and Thirring) with broader implications (Section~\ref{sec:discussion}). Technical details, including the off-shell Bethe Ansatz solution of the qKZ equations, are presented in Appendix~\ref{appendix}.

\section{General solution of the time-dependent Schr\"odinger equation}
\label{sec:SEsolution}

Here, we construct the general solution of the nonstationary Schr\"odinger equation, i.e., the solution to the partial differential equation
\bea
i\partial_t \ket{\Psi_N} = H(t) \ket{\Psi_N}, 
\label{SEk}
\eea
prior to imposing boundary conditions. Here, $N$ is the number of electrons and $H(t)$ is the Kondo Hamiltonian~\re{Hamiltonian} with a time-dependent spin-exchange coupling. Explicitly,
\beg
H(t) = -i \! \int {\hat \psi}^{\dagger}_{s}(x)\, \partial_x  {\hat \psi}_{s}(x)\, dx 
+  J(t)\,  {\hat \psi}^{\dagger}_{s}(0)\, \vec{\sigma}_{ss'} {\hat \psi}_{s'}(0) \cdot \vec{S}.
\label{Hamiltonian1}
\en
The functional form of $J(t)$ will be specified later, after we impose boundary conditions, and we suppress the limits of  integration over the spatial coordinate $x$ from now on.   The construction in this section closely follows Ref.~\cite{Parmesh1}, though we provide additional detail and highlight the physical meaning of certain steps.

As we shall see, the general solution contains one unconstrained amplitude. All other amplitudes in the wavefunction are related to this free amplitude through particle-particle and particle-impurity $S$-matrices, which we determine below. These relations encode the structure of scattering in the model and serve as the foundation for implementing boundary conditions in the next section.

 Our approach also parallels the standard Bethe Ansatz~\cite{Andrei1995}, but with one crucial difference.
 To highlight this distinction, it suffices to consider the case of a single electron interacting with the impurity.
In both frameworks---our time-dependent case and the conventional Bethe Ansatz---we begin by solving the free-particle Schr\"odinger equation away from the impurity:
\beg
i(\partial_t + \partial_x) f(x,t) = 0.
\label{1el_td_Sch}
\en
Here we have suppressed the spin indices in $f(x,t)$ for clarity.  
The   general solution of this equation is an arbitrary function of $x - t$, i.e., $f(x,t) = f(x - t)$. In the conventional   Bethe Ansatz, however, one seeks energy eigenstates,\footnote{The conventional Bethe Ansatz can thus be viewed as an ``on-shell'' approach, in contrast to the ``off-shell'' Bethe Ansatz employed here.} which fixes the time dependence of the wavefunction to be $f(x,t) \propto e^{-iEt}$. The requirement that $f$ depend only on $x - t$ then enforces the form $f(x,t) = C e^{iE(x - t)}$. That is, the combination of linear dispersion and stationarity constrains  the wavefunction to a plane wave, $e^{ikx}$, with $k = E$.

In contrast, we solve the full time-dependent Schr\"odinger equation, where no such restriction arises. The solution is a general function of $x - t$, whose form is fixed by boundary conditions rather than the requirement of stationarity.  This feature---the use of arbitrary functions of $x - t$ in place of plane waves---is a basic structural difference that will play a role throughout the paper.

The generalization to multiple particles is entirely analogous. For instance, the first-quantized Schr\"odinger equation for two free electrons reads:
\beg
i(\partial_t + \partial_{x_1} + \partial_{x_2}) f(x_1, x_2, t) = 0,
\en
with general solution $f(x_1, x_2, t) = f(x_1 - t, x_2 - t)$---an arbitrary function of both $x_1 - t$ and $x_2 - t$.

 Aside from this point, the structure of the general solution of the Schr\"odinger equation mirrors the standard Bethe Ansatz. The wavefunction consists of amplitudes associated with different orderings of particles in configuration space, each containing a spin part and a phase part.   The spin part of any amplitude can be obtained from a chosen reference amplitude by successive applications of the particle-particle and particle-impurity scattering matrices. Consequently, for any number of particles $N$, the full wavefunction is determined by a single unconstrained amplitude. We now carry out this construction explicitly, starting with the one-particle case.  
 
\subsection{One-particle solution}
\label{one_part_Schr}

A one-electron state takes the form
\bea
\label{1pformkb}
\ket{\Psi_1} =   \int  dx\; \hat{\psi}^{\dagger}_{s_1}(x) F_{s_1s_0}(x,t)\ket{s_0},
\eea
where $s_1$ and $s_0$ label the spin states of the electron and the impurity, respectively. Substituting this form into the time-dependent Schr\"odinger equation~(\ref{SEk}) yields
\beg
-i(\partial_t + \partial_x) F_{s_1 s_0}(x,t) + \delta(x)\left(J(t)\, \vec{\sigma}_{s_1s'_1} \cdot \vec{S}_{s_0s'_0} \right) F_{s'_1s'_0}(x,t) = 0.
\label{1pse}
\en
The second term, proportional to the Dirac delta function, represents the localized interaction between the electron and the impurity at $x = 0$.

Away from the impurity, i.e., for $x \ne 0$, the interaction term vanishes and Eq.~\re{1pse} reduces to
\beg
-i(\partial_t + \partial_x) F_{s_1 s_0}(x,t) = 0.
\label{1pse_free}
\en
As discussed earlier, the general solution of this equation is an arbitrary function of $x - t$. Since space is divided into two disjoint regions---to the left of the impurity ($x < 0$) and to the right ($x > 0$)---the solution generally takes different forms in each region:
\beg
F_{s_1 s_0}(x,t) = f^{10}_{s_1 s_0}(x - t)\, \theta(-x) + f^{01}_{s_1 s_0}(x - t)\, \theta(x).
\label{1pformk}
\en
Here, $f^{10}_{s_1 s_0}$ and $f^{01}_{s_1 s_0}$ are the amplitudes for the electron being to the left and right of the impurity, respectively. The superscripts indicate the spatial ordering relative to the impurity: `1' denotes the electron, and `0' the impurity. 
The Heaviside step function $\theta(x)$ is defined as
\be
\theta(x) = 
\begin{cases}
1, & x> 0, \\
\frac{1}{2}, & x=0,\\
0, & x < 0.
\end{cases}
\ee
Physically, the passage of the electron through the impurity at $x = 0$ induces a spin-dependent phase shift, encoded in the relation between $f^{10}$ and $f^{01}$.

The delta function term in Eq.~\re{1pse} imposes a boundary condition that connects the solutions on either side of the impurity. Substituting the expression~\re{1pformk} into the Schr\"odinger equation~\re{SEk}, we obtain the matching condition:
\beg
\label{1prel1}
f^{01}_{s_1 s_0}(z) = S^{10}_{s_1s_0, s_1' s_0' }(z)\, f^{10}_{s_1' s_0'}(z),
\en
where we have introduced the shorthand $z = x - t$. The operator $S^{10}(z)$ is the particle-impurity scattering matrix, given by
\begin{align}
\label{smatk}
S^{10}(z) &= e^{i\phi(z)} \frac{i g(z)\, I^{10} + P^{10} }{i g(z) + 1},\\
\label{gJ} g(z) &= \frac{1 - \frac{3}{16} J^2(-z)}{J(-z)},\\
e^{i\phi(z)} &= \frac{1 - i J(-z) - \frac{3}{16} J^2(-z)}{1 - \tfrac{1}{2}i J(-z) + \frac{3}{16} J^2(-z)}.
\end{align}
The superscript `10' indicates that the matrix acts on the combined spin space of the particle (`1') and the impurity (`0'). 

Here, $I^{kl}$ and $P^{kl}$ are the spin identity and spin-exchange (permutation) operators acting on the combined spin spaces $k$ and $l$.  Their action on a vector $A$ with  components $A_{s_k s_l}$ in this spin space is
\be
   \bigl(I^{kl}  A\bigr)_{s_k s_l} = A_{s_k s_l},\qquad    \bigl(P^{kl}  A\bigr)_{s_k s_l} = A_{s_l s_k}.
   \label{iden_action}
\ee
Since spin indices $s_k$ and $s_l$ each take two values, $\uparrow$ and $\downarrow$, both $I^{kl}$ and $P^{kl}$ are $4\times 4$ matrices with components
\beg
I^{kl}_{s_ks_l,s_k' s_l'}=\delta_{s_k s_k'} \delta_{s_l s_l'},\qquad P^{kl}_{s_k s_l,s_k' s_l'}=\delta_{s_k s_l'} \delta_{s_l s_k'}.
\label{iden_comp}
\en
Thus, $S^{10}(z)$  is  a $4\times 4$ matrix-valued function of the variable $z$. Note also that for a time-independent spin-exchange coupling $J(t)=\text{const}$, the $S$-matrix~\re{smatk} coincides with the one in the standard (on-shell) Bethe Ansatz for the Kondo Hamiltonian~~\cite{Andrei1995}.

In Eq.~\re{1pformk}, the amplitude $f^{10}(z)$ is associated with the physical region $x < 0$ and is unphysical for $x > 0$, while $f^{01}(z)$ is physical for $x > 0$ and unphysical for $x < 0$. The relation~\re{1prel1} thus allows one to express the physical amplitude on one side of the impurity in terms of the unphysical amplitude on the other side, mediated by the scattering matrix $S^{10}(z)$. Specifically, for $x > 0$, the physical amplitude $f^{01}(z)$ is obtained from $f^{10}(z)$ via $S^{10}(z)$, while for $x < 0$, $f^{10}(z)$ is related to $f^{01}(z)$ through the inverse matrix $\left[S^{10}(z)\right]^{-1}$.

At this stage, we see that the Hamiltonian imposes a constraint relating the two amplitudes, leaving one of them free. To fully determine the solution in the one-particle sector, we impose periodic boundary conditions. This yields a quantization condition on $f^{10}(z)$, and equivalently on $f^{01}(z)$ via Eq.~\re{1prel1}. We will show later that this condition takes the form of a functional (difference) equation, whose solution determines the explicit form of $f^{10}(z)$ in terms of the interaction strength $J(t)$.

\subsection{Two-particle solution}

We now turn to the two-particle sector. Two-particle states are of the form
\be
\label{2pformk}
\ket{\Psi_2} =  \iint dx_1\, dx_2\; \hat{\psi}^{\dagger}_{s_1}(x_1) \hat{\psi}^{\dagger}_{s_2}(x_2)\, \mathcal{A} F_{s_1s_2s_0}(x_1,x_2,t)\ket{s_0},
\ee
where $s_1$ and $s_2$ are the spin indices of the electrons, and $s_0$ labels the spin state of the impurity. The operator $\mathcal{A}$ denotes antisymmetrization under exchange of both  electron positions and spins, i.e., under $x_1 \leftrightarrow x_2$ and $s_1 \leftrightarrow s_2$.

Substituting Eq.~\re{2pformk} into the Schr\"odinger equation~\re{SEk}, we obtain
\beg
\label{2pdiffk}
-i(\partial_t + \partial_{x_1} + \partial_{x_2}) F_{s_1 s_2s_0} 
+ J(t)\Big[ \delta(x_1)\, \vec{\sigma}_{s_1 s_1'} \cdot \vec{S}_{s_0 s_0'}\, \delta_{s_2 s_2'}
+ \delta(x_2)\, \vec{\sigma}_{s_2 s_2'} \cdot \vec{S}_{ s_0 s_0'}\, \delta_{s_1 s_1'} \Big] F_{s_1' s_2' s_0'} = 0.
\en

As in the one-particle case, the amplitude depends on the position of each 
particle relative to the impurity. In addition, we choose to distinguish 
amplitudes corresponding to different ordering of the particles with respect 
to one another. Since the particles move with the same (Fermi) velocity, their 
relative ordering is conserved by the Kondo Hamiltonian.

With two particles, we thus consider six distinct spatial orderings:
\begin{enumerate}
\item Both particles on the left side of the impurity, with particle 1 to the 
left of 2\\ 
($x_1 < x_2 < 0$).
\item Both particles on the left side of the impurity, with particle 2 to the 
left of 1\\
 ($x_2 < x_1 < 0$).
\item Both particles on the right, with particle 1 to the right of 2 
($x_1 > x_2 > 0$).
\item Both particles on the right, with particle 2 to the right of 1 
($x_2 > x_1 > 0$).
\item Particle 1 on the left and particle 2 on the right ($x_1 < 0$, 
$x_2 > 0$).
\item Particle 2 on the left and particle 1 on the right ($x_2 < 0$, 
$x_1 > 0$).
\end{enumerate}

Each configuration corresponds to a distinct amplitude in the wavefunction. 
Each amplitude solves the free two-particle time-dependent Schr\"odinger equation 
in the corresponding region of the two-dimensional $(x_1, x_2)$ space---that is, 
it is a function of $z_1 = x_1 - t$ and $z_2 = x_2 - t$. We therefore write:
\beg
\begin{split}
  F(x_1,x_2,t)&= f^{120}(z_1,z_2)\:\theta(-x_2)\theta(-x_1) \theta(x_2-x_1)+
  f^{210}(z_1,z_2)\:\theta(-x_2)\theta(-x_1)\theta(x_1-x_2)\\
   &+f^{012}(z_1,z_2)\:\theta(x_1)\theta(x_2)\theta(x_2-x_1)+f^{021}(z_1,z_2)\:\theta(x_1)\theta(x_2)\theta(x_1-x_2)\\
&+f^{102}(z_1,z_2)\:\theta(-x_1)\theta(x_2)+f^{201}(z_1,z_2)\:\theta(-x_2)\theta(x_1).
\end{split}
\label{2pwfk}
\en  

As in the one-particle sector, the amplitudes, where a particle is to immediate left and immediate right of the impurity, are related to each other through  the particle-impurity $S$-matrix,
\beg
 \begin{split} 
f^{201}(z_1,z_2)=S^{10}(z_1)f^{210}(z_1,z_2), \qquad f^{102}(z_1,z_2)=S^{20}(z_2)f^{120}(z_1,z_2),\\
f^{012}(z_1,z_2)=S^{10}(z_1)f^{102}(z_1,z_2), \qquad f^{021}(z_1,z_2)=S^{20}(z_2)f^{201}(z_1,z_2).
\end{split}
\label{2psmatk}
\en
Here and for the rest of this section we suppress spin indices on the amplitudes and scattering matrices.

We introduced a redundancy when we split the free two-particle solution on either 
side of the impurity into two parts---one where particle 1 is to the left of particle 2, 
and the other where it is to the right. In this representation, a single arbitrary function 
is written in terms of two independent arbitrary functions.  Consider, for example, the case 
when both particles are to the left of the impurity. The general solution for each spin 
component is an arbitrary function of $z_1$ and $z_2$, yet we expressed it in terms 
of two functions, $f^{120}(z_1, z_2)$ and $f^{210}(z_1, z_2)$, as
\be
f^{<0}(z_1, z_2) = f^{120}(z_1, z_2)\: \theta(z_2 - z_1) + 
f^{210}(z_1, z_2)\: \theta(z_1 - z_2),
\label{<0}
\ee
where we have used $x_2 - x_1 = z_2 - z_1$.

As a result, we are free to impose a constraint between $f^{120}$ and $f^{210}$ 
without loss of generality. We relate them using a suitably chosen particle-particle 
$S$-matrix. The same applies to the analogous pair of amplitudes corresponding to 
both particles being to the right of the impurity:
\be
f^{210}(z_1, z_2) = S^{12}(z_1, z_2)\, f^{120}(z_1, z_2), \qquad
f^{021}(z_1, z_2) = S^{12}(z_1, z_2)\, f^{012}(z_1, z_2),
\label{eemateqs}
\ee
where
\be
S^{12}(z_1, z_2) = 
\frac{i [g(z_1) - g(z_2)]\, I^{12} + P^{12}}{i [g(z_1) - g(z_2)] + 1}.
\label{smatee}
\ee

Indeed, any function $f^{<0}(z_1, z_2)$ can be expressed in the form~\re{<0} with the amplitudes $f^{120}$ and $f^{210}$ constrained by Eq.~\re{eemateqs}, by defining
 \be
f^{120}(z_1, z_2) =
\begin{cases}
f^{<0}(z_1, z_2), & z_2 \ge z_1, \\
\left[S^{12}(z_1, z_2)\right]^{-1} f^{<0}(z_1, z_2), & z_2 < z_1.
\end{cases}
\ee

We choose the form~\re{smatee} so that the electron-electron   (\ref{smatee}) and  electron-impurity  (\ref{2psmatk}) $S$-matrices satisfy the Yang-Baxter algebra,
\be 
S^{20}(z_2)S^{10}(z_1)S^{12}(z_1,z_2)=S^{12}(z_1,z_2)S^{10}(z_1)S^{20}(z_2).
\ee
Note that an analogous choice of the arbitrary electron-electron scattering matrix is made in the on-shell Bethe Ansatz~\cite{Andrei1995}. Moreover,   for a time-independent coupling,  this   choice coincides with the $S$-matrix~\re{smatee}, since  $g(z)=\mathrm{const}$ in this case.

The relations~\re{2psmatk} and~\re{eemateqs} allow us to express all amplitudes in the two-particle wavefunction~\re{2pwfk} in terms of any one amplitude of our choosing. As in the one-particle case, this means that---prior to imposing boundary conditions---there is a single free amplitude. We will see later that applying boundary conditions leads to a constraint equation for this free amplitude, which takes the form of a matrix difference equation. Solving this equation yields an explicit expression for the free amplitude. The remaining amplitudes in the wavefunction~\re{2pwfk} can then be reconstructed using the relations~\re{2psmatk} and~\re{eemateqs}.

\subsection{$N$-particle solution}
\label{Sec:N_particle}

A general $N$-particle state is given by
\begin{equation}
\label{npform1}
\ket{\Psi_N} =   \prod_{j=1}^{N} \int  dx_j\, \hat{\psi}^{\dagger}_{s_j}(x_j)\, \mathcal{A} F_{s_0 \dots s_N }(x_1, \dots, x_N, t)\ket{s_0},
\end{equation}
where $s_j$ denote the spin indices of the electrons and the impurity, and $\mathcal{A}$ denotes antisymmetrization over both coordinates $x_j$ and spin indices $s_j$ of the electrons.
The wavefunction $F$ is constructed as a sum over amplitudes corresponding to different orderings of the particle and impurity coordinates:
\begin{equation}
\label{npwf_general}
F_{ s_0 \dots s_N}(x_1, \dots, x_N, t)
= \sum_Q \theta\left(x_Q\right)\, f^Q_{ s_0 \dots s_N}(z_1, \dots, z_N),
\end{equation}
with $z_j = x_j - t$ and
\be
\theta\left(x_Q\right) = 
\begin{cases}
1, & x_{Q(0)} < \dots < x_{Q(N)}, \\
\frac{1}{2},& \text{if $x_{Q(i)}=x_{Q(j)}$ for any pair of $i\ne j$},\\
0, & \text{otherwise}.
\end{cases}
\ee
 The sum runs over all permutations $Q$ of particle labels, and each amplitude $f^Q$ solves the free Schr\"odinger equation in the corresponding sector of configuration space. The impurity coordinate is fixed at $x_0=0$.

As in the two-particle case, amplitudes that differ by exchanging the order of two particles are related by a particle-particle $S$-matrix:
\begin{equation}
f^{\dots ji \dots}(z_1,\dots,z_N) = S^{ij}(z_i, z_j)\, f^{\dots ij \dots}(z_1,\dots,z_N),
\label{eemateqsn}
\end{equation}
where
\begin{equation}
S^{ij}(z_i,z_j) = \frac{i[g(z_i) - g(z_j)]\, I^{ij} + P^{ij}}{i[g(z_i) - g(z_j)] + 1},
\label{smateen}
\end{equation}
and $I^{ij}$ and $P^{ij}$ are the spin-identity and spin-exchange operators acting on particles $i$ and $j$. These matrices satisfy the Yang-Baxter algebra:
\begin{equation}
S^{ij}(z_i,z_j)\, S^{ik}(z_i,z_k)\, S^{jk}(z_j,z_k)
= S^{jk}(z_j,z_k)\, S^{ik}(z_i,z_k)\, S^{ij}(z_i,z_j).
\label{YB3}
\end{equation}

Similarly, the amplitudes in which a particle $j$ is to the left or right of the impurity are related by the particle-impurity $S$-matrix:
\begin{equation}
f^{\dots 0j \dots}(z_1,\dots,z_N) = S^{j0}(z_j)\, f^{\dots j0 \dots}(z_1,\dots,z_N),
\label{1prel}
\end{equation}
with
\begin{align}
S^{j0}(z) &= e^{i\phi(z)}\, \frac{i g(z)\, I^{j0} + P^{j0}}{i g(z) + 1},\label{sj0} \\
g(z) &= \frac{1 - \tfrac{3}{16} J^2(-z)}{J(-z)}, \\
\label{phase} e^{i\phi(z)} &= \frac{1 - i J(-z) - \tfrac{3}{16} J^2(-z)}{1 - \tfrac{1}{2}i J(-z) + \tfrac{3}{16} J^2(-z)}.
\end{align}
The particle-particle and particle-impurity $S$-matrices satisfy the mixed Yang-Baxter relation:
\begin{equation}
S^{j0}(z_j)\, S^{i0}(z_i)\, S^{ij}(z_i, z_j) = S^{ij}(z_i, z_j)\, S^{i0}(z_i)\, S^{j0}(z_j).
\label{YB2}
\end{equation}

These relations ensure that all amplitudes in the $N$-particle wavefunction can be obtained from a single reference amplitude by successive applications of $S$-matrices. That is, before imposing boundary conditions, there exists one unconstrained amplitude in the wavefunction. This mirrors the structure found in the one- and two-particle cases and reflects the integrability of the model.

In summary, the general solution of the time-dependent Schr\"odinger equation is 
built from amplitudes defined in regions of configuration space labeled by particle 
orderings. All amplitudes are related by particle-particle and particle-impurity 
$S$-matrices. In the next section, we impose periodic boundary conditions, which 
constrain the previously free amplitude through a set of matrix difference 
equations. Solving these yields the full wavefunction via the $S$-matrix relations.

\section{Periodic boundary conditions}

We now consider the system on a finite interval,
\beg
-\tfrac{L}{2} \le x \le \tfrac{L}{2},
\en
of length~$L$, and impose periodic boundary conditions on the wavefunction 
$F_{s_0 \dots s_N }(x_1, \dots, x_N, t)$. 
As we shall see, this introduces constraint equations on the previously free amplitude. 
These take the form of matrix difference equations, which must be solved to obtain the explicit form of the amplitude. 
In the next section, we map these equations to the \textsl{quantum Knizhnik--Zamolodchikov equations}.
The remaining amplitudes can then be recovered by applying the $S$-matrices established earlier.

Explicitly, periodic boundary conditions read
\be 
\label{pbc}
\begin{split}
F\left(-\tfrac{L}{2}, x_2, \dots, x_N, t\right)
&= F\left(\tfrac{L}{2}, x_2, \dots, x_N, t\right), \\
F\left(x_1, -\tfrac{L}{2}, \dots, x_N, t\right)
&= F\left(x_1, \tfrac{L}{2}, \dots, x_N, t\right), \\
\vdots \qquad  \qquad \; &\qquad \qquad \; \vdots \\
F\left(x_1, x_2, \dots, -\tfrac{L}{2}, t\right)
&= F\left(x_1, x_2, \dots, \tfrac{L}{2}, t\right).
\end{split}
\ee
That is, we identify the endpoints $-\frac{L}{2}$ and $\frac{L}{2}$ of the interval so that the particles move on a ring. 
For notational simplicity, we omit spin indices in what follows.

Recall that the wavefunction is decomposed into sectors labeled by distinct orderings $Q$ of particle positions as
\begin{equation}
\label{npwf_pbc}
F(x_1, \dots, x_N, t)
= \sum_Q \theta\left(x_Q \right)\,
f^Q(x_1 - t, \dots, x_N - t).
\end{equation}
A particle $j$ can appear at position $x_j = -\frac{L}{2}$ only when it is the leftmost particle on the left of the impurity. 
This corresponds to amplitudes of the form $f^{j\dots0\dots}$. 
Likewise, it can appear at $x_j = \frac{L}{2}$ only when it is the rightmost particle on the right of the impurity,
corresponding to amplitudes $f^{\dots0\dots j}$. 
Accordingly, the $j$-th boundary condition in~\re{pbc} imposes the constraint
\beg
f^{j\dots0\dots}\Bigl(x_1 - t, \dots, 
\underset{\text{particle } j}{-\tfrac{L}{2} - t}, 
\dots, x_N - t\Bigr)
= f^{\dots0\dots j}\Bigl(x_1 - t, \dots, 
\underset{\text{particle } j}{\tfrac{L}{2} - t}, 
\dots, x_N - t\Bigr),
\label{x-tconstraint}
\en
where the ordering of all particles except particle \( j \) is the same on both sides.

Since \( \tfrac{L}{2} - t \) can take arbitrary real values \( z_j \), 
these constraints are equivalent to
\beg
f^{j\dots0\dots}\left(z_1, \dots,  z_j - L, \dots, z_N \right)
= f^{\dots0\dots j}\left(z_1, \dots,  z_j, \dots, z_N\right),\quad j=1,\dots, N.
\label{periodic}
\en
To illustrate the structure of these constraints, we proceed sector by sector: 
first for a single particle, then for two particles, and finally for a general $N$-particle state.

\subsection{Single-particle sector}

In the single-electron case, there is only one  periodicity constraint from Eq.~\re{periodic}, namely,
\be
f^{10} _{s_1s_0}(z - L) = f^{01} _{s_1 s_0}(z).
\label{1pbck}
\ee
Here, $f^{10}$ and $f^{01}$ denote the amplitudes for the particle being to the left and to the right of the impurity, respectively.

As shown in Sec.~\ref{one_part_Schr}, the time-dependent Schr\"odinger equation implies a relation between these amplitudes via the particle-impurity $S$-matrix:
\be
\label{1prel1111}
f^{01} _{s_1 s_0}(z) = S^{10} _{s_1s_0, s_1's_0'}(z)\, f^{10}_{s_1' s_0'}(z),
\ee
with $S^{10}(z)$ given explicitly in Eq.~\re{smatk}.
Substituting this into the periodicity condition \re{1pbck}, we obtain a matrix difference equation for $f^{10}(z)$:
\be
\label{1pdiffk}
f^{10} _{s_1 s_0}(z - L) = S^{10} _{s_1s_0, s_1's_0'}(z)\, f^{10}_{s_1' s_0'}(z).
\ee
This equation constrains the form of the amplitude $f^{10}(z)$ in terms of the interaction strength $J(t)$ encoded in the $S$-matrix. It is the first in a sequence of such difference equations that we will derive in the multi-particle sectors below.

 Before proceeding further, let us compare this with the on-shell (standard) Bethe Ansatz~\cite{Andrei1995} for time-independent interaction strength, $J(t) = J$. 
As discussed below Eq.~\re{1el_td_Sch}, in this case we seek stationary states of the form $f \propto e^{-iEt}$. 
Due to the linear dispersion relation, this fixes the time and coordinate dependence of the amplitudes to be
\be
\begin{split}
f^{10} _{s_1 s_0}(z) &= A^{10} _{s_1 s_0}\, e^{ikz} = A^{10} _{s_1 s_0}\, e^{ik(x - t)}, \\
f^{01} _{s_1 s_0}(z) &= A^{01} _{s_1 s_0}\, e^{ikz} = A^{01} _{s_1 s_0}\, e^{ik(x - t)},
\end{split}
\ee
where $k = E$ (since $v_F = 1$) is the single-particle momentum, which also equals the energy.
The difference equation~\re{1pdiffk} now becomes a simple eigenvalue equation for the particle-impurity $S$-matrix $S^{10} _{s_1s_0, s_1's_0'}$:
\be
e^{-ikL}\, A^{10} _{s_1 s_0} = S^{10} _{s_1s_0, s_1's_0'}\, A^{10}_{s_1's_0'}.
\ee
Here, the $S$-matrix is time-independent and given by Eq.~\re{smatk} with $J(-z) = J$.

In contrast, for time-dependent interaction strength, Eq.~\re{1pdiffk} remains a matrix difference equation that must be solved to determine the explicit functional form of the amplitude $f^{10} _{s_1 s_0}(z)$. Once this amplitude is known, Eq.~\re{1pbck} yields the corresponding amplitude $f^{01} _{s_1 s_0}(z)$, thereby reconstructing the full one-particle wavefunction~\re{1pformk}.  

\subsection{Two-particle sector}

Periodic boundary conditions~\re{periodic} specialized to the two-electron case read (suppressing spin indices)
\beg
\begin{split}
f^{201}(z_1, z_2) &= f^{120}(z_1 - L, z_2), \qquad 
f^{102}(z_1, z_2) = f^{210}(z_1, z_2 - L), \\
f^{021}(z_1, z_2) &= f^{102}(z_1 - L, z_2), \qquad 
f^{012}(z_1, z_2) = f^{201}(z_1, z_2 - L).
\end{split}
\label{2pbck}
\en
Using these equations in combination with the $S$-matrix relations~\re{2psmatk} and~\re{eemateqs}, we obtain
\begin{align}
f^{210}(z_1 - L, z_2) &= S^{12}(z_1, z_2 + L)\, S^{10}(z_1)\, f^{210}(z_1, z_2),
\label{2pdiffeqk3} \\
f^{210}(z_1, z_2 - L) &= S^{20}(z_2)\, S^{12}(z_2, z_1)\, f^{210}(z_1, z_2).
\label{2pdiffeqk4}
\end{align}
Equations~\re{2pdiffeqk3} and~\re{2pdiffeqk4} respectively describe transporting particle~1 and particle~2 once around the ring. 
They form a system of matrix difference equations for the amplitude $f^{210}(z_1, z_2)$.
An analogous system exists for the amplitude $f^{120}(z_1, z_2)$.

Solving these equations determines the explicit form of $f^{210}(z_1, z_2)$. 
All other amplitudes in the two-particle wavefunction can then be reconstructed using the $S$-matrix relations~\re{2psmatk} and~\re{eemateqs}.

 \subsection{$N$-particle sector}

Consider the constraint imposed by the periodic boundary conditions~\re{periodic} on the amplitude 
$f^{N\,N{-}1\dots10}_{ s_0\dots s_N}$:
\beg
f^{N\,N{-}1\dots10}\left(z_1, \dots, z_{N-1}, z_N - L\right)
= f^{N{-}1\dots10\,N}\left(z_1, \dots, z_{N-1}, z_N\right).
\label{periodic111}
\en
This amplitude corresponds to the ordering $x_N < x_{N-1} < \dots < x_1 < 0$, where particle $N$ is 
the leftmost particle. On the right-hand side of \eref{periodic111}, particle $N$ has been transported 
around the ring, becoming the rightmost particle, while the relative ordering of all other particles 
remains unchanged.

We now construct an operator $Z_N$ that transports particle $N$ across the system, using a sequence 
of particle-particle and particle-impurity $S$-matrices:
\beg 
Z_N(z_1, \dots, z_N) = S^{N0}(z_N)\, S^{N1}(z_N, z_1)\, \cdots\, S^{N\,N{-}1}(z_N, z_{N-1}).
\label{transfermatN}
\en
This operator relates the two amplitudes in~\re{periodic111} as
\be
f^{N{-}1\dots10\,N}_{ s_0\dots s_N}(z_1, \dots, z_N)
= Z_N(z_1, \dots, z_N)\, f^{N\,N{-}1\dots10}_{ s_0\dots s_N}(z_1, \dots, z_N).
\label{shiftN}
\ee
Combining~\re{periodic111} and~\re{shiftN}, we obtain the matrix difference equation
\be
f^{N\,N{-}1\dots10}_{ s_0\dots s_N}(z_1, \dots, z_N - L)
= Z_N(z_1, \dots, z_N)\, f^{N\,N{-}1\dots10}_{ s_0\dots s_N}(z_1, \dots, z_N).
\label{diffeqtN}
\ee

More generally, we define an operator $Z_j$ that transports particle $j$ once around the ring, yielding
\be
f^{N\dots j \dots10}_{  s_0\dots s_N}(z_1, \dots, z_j - L, \dots, z_N)
= Z_j(z_1, \dots, z_N)\, f^{N\dots j \dots10}_{  s_0\dots s_N}(z_1, \dots, z_j, \dots, z_N),
\label{GqKZ}
\ee
where
\begin{equation}
\begin{split}
Z_j(z_1, \dots, z_N) 
&= S^{j\,j{+}1}(z_j, z_{j+1} + L)\, S^{j\,j{+}2}(z_j, z_{j+2} + L)\cdots S^{jN}(z_j, z_N + L) \\
&\quad \times S^{j0}(z_j)\, S^{j1}(z_j, z_1)\cdots S^{j\,j{-}1}(z_j, z_{j-1}).
\end{split}
\label{transfermat}
\end{equation}
There are thus $N$ such transport operators $Z_j$, $j = 1, \dots, N$, each acting on the amplitude 
$f^{N\dots10}_{  s_0\dots s_N}(z_1, \dots, z_N)$. Similar difference equations hold for the rest of the amplitudes in the $N$-particle wavefunction.

For consistency, transporting particle $j$ around the system followed by particle $i$ must yield the 
same result as transporting them in the opposite order. This requires
\be
Z_i(z_1, \dots, z_j - L, \dots, z_N)\, Z_j(z_1, \dots, z_N)
= Z_j(z_1, \dots, z_i - L, \dots, z_N)\, Z_i(z_1, \dots, z_N).
\label{commutations}
\ee
Since the transport operators $Z_j$ are built from particle-particle and particle-impurity $S$-matrices, 
these commutation relations impose consistency conditions on the $S$-matrices. Because the $S$-matrices 
depend on the interaction strength $J(t)$, equations~\re{commutations} in turn constrain the allowed 
time dependence of $J(t)$. Hence, these can be interpreted as integrability conditions.

To summarize, we derived an $N$-particle wavefunction~\re{npform1} that solves the time-dependent 
Schr\"odinger equation, provided that amplitudes corresponding to different particle orderings are 
related via the $S$-matrices~\re{smateen} and~\re{sj0}. Using these relations, all amplitudes 
can be written in terms of a single reference amplitude 
$f^{N\dots j \dots10}_{  s_0\dots s_N}(z_1, \dots, z_j, \dots, z_N)$, which is initially free. 
We further constructed an operator that transports a particle around the ring, and showed that periodic 
boundary conditions impose a matrix difference equation on this reference amplitude. 
If $J(t)$ satisfies the integrability condition~\re{commutations}, this equation can be solved 
explicitly, as we will see in the next section. The full wavefunction can then be recovered using the known $S$-matrix relations. 
Thus, finding the many-body wavefunction reduces to solving a set of matrix difference equations.

 Let us compare this procedure with the standard Bethe Ansatz approach applied when the interaction strength is constant. In the case $J(t) = J$, the $S$-matrices~\re{smatk} and~\re{smateen}, and consequently the transport operators,
\beg
Z_j = S^{j\,j{+}1} \cdots S^{jN}\, S^{j0}\, S^{j1} \cdots S^{j\,j{-}1},
\label{eigenN}
\en
become independent of $z$  and  $Z_j$ turns into standard monodromy matrix \cite{Andrei80,Andrei1995}.
The integrability conditions~\re{commutations} now reduce to simple commutativity requirements:
\beg
[Z_i, Z_j] = 0,
\en
which are satisfied as a consequence of the Yang-Baxter equations~\re{YB3} and~\re{YB2}.

Moreover, the stationarity of the wavefunction enforces a plane wave form for the amplitudes:
\beg
f^{N\dots10}_{  s_0\dots s_N}(z_1, \dots, z_N)
= A^{N\dots10}_{  s_0\dots s_N} \prod_{j=1}^N e^{i k_j z_j},
\en
where $A^{N\dots10}_{  s_0\dots s_N}$ are $z$-independent coefficients.

Substituting this into the difference equation~\re{GqKZ}, we find that it turns into an eigenvalue problem:
\beg
e^{i k_j L}\, A^{N\dots10}_{  s_0\dots s_N}
= Z_j\, A^{N\dots10}_{  s_0\dots s_N}.
\en
The operator~\re{eigenN} is then diagonalized using the standard algebraic Bethe Ansatz,
yielding the amplitudes $A^{N\dots10}_{  s_0\dots s_N}$~\cite{Andrei1995}.

\section{Mapping to quantum Knizhnik-Zamolodchikov equation}
\label{sec:mapping_qkz}

In previous sections, we reduced the problem of solving the nonstationary Schr\"odinger equation for the Kondo Hamiltonian with time-dependent spin-exchange coupling $J(t)$ to a system of matrix difference equations. Here, we show that for certain choices of the time dependence of $J(t)$, this system maps to the 
\textsl{quantum Knizhnik--Zamolodchikov  equations (qKZ)}~\cite{Frenkel,rishetikhin1,smirnovqKZ,smirnovqkz2,JimboqKZ,Babujian_1997,Etingof1998} 
associated with the ${su}(2)$ Lie algebra. This connection enables us to leverage the integrability of the qKZ system to construct the many-body wavefunction of the time-dependent Kondo Hamiltonian.

The original Knizhnik--Zamolodchikov (KZ) equations arose in conformal field theory (CFT) as differential equations governing correlation functions in Wess--Zumino--Witten  models~\cite{KnizhnikZamolodchikov1984}. They describe how conformal blocks depend on the insertion points of primary fields and encode the analytic continuation and braiding properties of quantum amplitudes in two-dimensional CFTs with affine Lie algebra symmetry.

A quantum deformation of the KZ equations, known as the \textsl{quantum} KZ (qKZ) equations, emerged in the study of quantum integrable models and quantum groups. These are \textsl{difference} equations, as opposed to differential ones, and govern how form factors or correlation functions change under discrete shifts in particle positions. In the appropriate scaling limit---when both the step size and the crossing parameter tend to zero---the qKZ equations reduce to the original KZ equations~\cite{Frenkel,JimboqKZ,Etingof1998,Nakayashiki1999}.

In our setting, the qKZ equations arise as consistency conditions imposed by periodic boundary conditions on the amplitudes of the many-body wavefunction. For specific time dependences of the interaction $J(t)$, the resulting matrix difference equations reduce to qKZ equations associated with the $su(2)$ Lie algebra. 

The quantum Knizhnik--Zamolodchikov equations are a system of matrix  difference equations:
\be
\varphi(y_0, \dots, y_j + \kappa,\dots, y_N) 
= M_j(y_0, y_1,\dots,y_N)\, \varphi(y_0,\dots,y_j,\dots,y_N),
\label{diffeq12}
\en
where $\varphi(y_0, \dots, y_N)$ is a vector-valued function with components 
$\varphi_{s_0 \dots s_N}(y_0, \dots, y_N)$, 
and  $s_0,\dots, s_N$ are spin-$\frac{1}{2}$ labels taking values $\uparrow$ and $\downarrow$. The real parameter $\kappa$ is referred to as the \textsl{step}. The matrix $M_j(y_0,\dots,y_N)$ is constructed from the XXX $R$-matrices $R^{ij}(y)$ as
\begin{align}
M_j(y_0,\dots,y_N) &= 
R^{j+1\,j}(y_{j+1} - y_j - \kappa)\, R^{j+2\,j}(y_{j+2} - y_j - \kappa) \dots R^{N\,j}(y_N - y_j - \kappa)
\nonumber\\
&\quad \times R^{0\,j}(y_0-y_j)\, R^{1\,j}(y_1 - y_j) \dots R^{j-1\,j}(y_{j-1} - y_j),
\label{qKZ}
\end{align}
 and satisfies the following relations
 \begin{align}M_j(y_0,...,y_k+\kappa,...,y_N)M_k(y_0,...,y_N)=M_k(y_0,...,y_j+\kappa,...y_N)M_j(y_0,...,y_N).
\end{align}
Each $R$-matrix $R^{ij}(y)$ is a $4 \times 4$ matrix acting on the spin degrees of freedom of particles $i$ and $j$, and is given by
\be
R^{ij}(y) = \frac{i y\, I^{ij} + c\, P^{ij}}{i y + c},
\label{rmatdef}
\en
where $c$ is the \textsl{crossing parameter}, and $I^{ij}$ and $P^{ij}$ are the identity and permutation operators acting on the combined spin space of particles $i$ and $j$. These operators were defined earlier in Eqs.~\re{iden_action} and~\re{iden_comp}.

Let us compare the qKZ equations with the matrix difference equations derived in the previous section [see~\eref{GqKZ}], which we reproduce here for convenience:
\be
f(z_1, \dots, z_j - L, \dots, z_N)
= Z_j(z_1, \dots, z_N)\, f(z_1, \dots, z_j, \dots, z_N),
\label{GqKZ1}
\ee
where the transport operator $Z_j(z_1, \dots, z_N)$ is given by
\begin{equation}
\begin{split}
Z_j(z_1, \dots, z_N) 
&= S^{j\,j{+}1}(z_j, z_{j+1} + L)\, S^{j\,j{+}2}(z_j, z_{j+2} + L) \cdots S^{jN}(z_j, z_N + L) \\
&\quad \times S^{j0}(z_j)\, S^{j1}(z_j, z_1) \cdots S^{j\,j{-}1}(z_j, z_{j-1}).
\end{split}
\label{transfermat1}
\end{equation}
We have suppressed the lower spin indices $ s_0 \dots s_N$ in~\eref{GqKZ1} to match the notation used in~\eref{diffeq12}, and we omit the fixed amplitude label $N \dots j \dots 10$, which is the same in all $N$ equations of the system~\re{GqKZ1}. Recall also the form 
of the particle-impurity and particle-particle scattering matrices [\esref{sj0} and \re{smateen}]:
\be
S^{j0}(z) = e^{i\phi(z)}\, \frac{i g(z)\, I^{j0} + P^{j0}}{i g(z) + 1},
\label{sj01}
\en
\begin{equation}
S^{ij}(z_i,z_j) = \frac{i[g(z_i) - g(z_j)]\, I^{ij} + P^{ij}}{i[g(z_i) - g(z_j)] + 1}.
\label{smateen1}
\end{equation}

We observe that the qKZ equations~\re{diffeq12} and the periodic boundary conditions~\re{GqKZ1} coincide provided the following conditions are met:
\begin{enumerate}

\item The phase factor $e^{i\phi(z_j)}$ entering $Z_j$ through the particle-impurity scattering matrix $S^{j0}(z_j)$ in~\eref{transfermat1} is eliminated.

\item The parameters in the qKZ system are chosen as
\be
\frac{y_i}{c} = -g(z_i), \qquad \frac{y_i - \kappa}{c} = -g(z_i + L), \qquad y_0 = 0,
\label{yg}
\en
for all $i \ge 1$.
\end{enumerate}
To remove the phase factor from the $S$-matrix, we redefine the amplitude as
\be
f(z_1, \dots, z_N) = \tilde{f}(z_1, \dots, z_N) \prod_{i=1}^N h(z_i),
\label{qKZ_con_2}
\en
where $h(z)$ is a scalar-valued function satisfying the functional equation
\be
h(z - L) = e^{i\phi(z)}\, h(z).
\label{heq}
\en
In addition, the transformation equations~\re{yg} are compatible if and only if the function \( g(z) \) satisfies
\be
g(z + L) = g(z) + \frac{\kappa}{c},
\label{geq}
\en
which constrains the functional form of $g(z)$, and thereby of the interaction strength $J(t)$, as will be determined below.

Given \esref{heq} and \re{geq}, the vector-valued function  
\begin{equation}  
\tilde{f}(z_1, \dots, z_N) = \varphi(0, -c\,g(z_1), \dots, -c\,g(z_N))  
\label{qKz_con_1}
\end{equation}  
satisfies the qKZ equations~\re{diffeq12}. Thus, the known solution of the qKZ equations determines $\tilde{f}(z_1, \dots, z_N)$, and thereby---by construction---the amplitude 
\beg
f^{N \dots 10}(z_1, \dots, z_N)\equiv f(z_1, \dots, z_N) 
\label{qKz_con_3}
\en
in the time-dependent Kondo wavefunction. The remaining amplitudes, and hence the full solution of the nonstationary Schr\"odinger equation, can then be obtained using the known $S$-matrix relations.  

Note that we do not use Eq.~\re{diffeq12} with $j = 0$, which corresponds to the impurity. This equation is redundant, as transporting the impurity around the system is equivalent to transporting all $N$ particles sequentially.

It remains to solve Eqs.~\eqref{heq} and~\eqref{geq}. The general solution of Eq.~\eqref{geq} is
\begin{equation}
g(z) = \frac{\kappa}{c L} z + \text{periodic}(z),
\label{periodic1}
\end{equation}
where $\text{periodic}(z)$ denotes an arbitrary function of period $L$.
Now recall the relation between the spin-exchange coupling $J(t)$ and the function $g(z)$ [see Eq.~\eqref{gJ}],
\begin{equation}
\frac{1 - \tfrac{3}{16} J^2(z)}{J(z)} = g(-z).
\label{gJ1}
\end{equation}
Solving this equation for $J(z)$, we find
\begin{equation}
J(t) = \lambda t + p(t) \pm \sqrt{[\lambda t + p(t)]^2 + \tfrac{16}{3}},
\label{Jgeneral}
\end{equation}
where $p(t)$ is an arbitrary periodic function of period $L$, and the parameter $\lambda$ sets the characteristic timescale over which the spin-exchange interaction varies. It is related to the step size $\kappa$, the crossing parameter $c$, and the system size $L$ via
\begin{equation}
\lambda = \frac{8\kappa}{3cL}.
\end{equation}
In particular, setting $p(t) \equiv 0$, which corresponds to a linear mapping $g(z) = \kappa z / (cL)$ to the qKZ variables, yields 
\begin{equation}
J(t) = \lambda t \pm \sqrt{(\lambda t)^2 + \tfrac{16}{3}}.
\label{Jexplicit}
\end{equation}
This provides a concrete example of a time-dependent spin-exchange coupling for which the nonstationary Schr\"odinger equation for the Kondo Hamiltonian is exactly solvable. Other examples can be generated by including a nonzero periodic part in Eq.~\eqref{Jgeneral}.  As mentioned before, in the limit $\lambda\rightarrow \infty$, the above time-dependent interaction strength reduces to $J(t)=C/(a+t)$, the case previously studied in \cite{Parmesh1}.
Finally, note that the coupling $J(t)$---and hence the time-dependent Kondo Hamiltonian and the resulting wavefunction---depends on the step size $\kappa$ and the crossing parameter $c$ only through their ratio $\kappa/c$. Therefore, without loss of generality, we may set $\kappa = 1$.

 Observe that we require only a particular solution of~\eref{heq} to perform the mapping to the qKZ equations. For real $J(-z)$, Eq.~\eqref{phase} gives the following continuous choice of the phase:
\begin{equation}
 \phi(z) = -2\arctan\left[\frac{J(-z)}{4}\right].
\end{equation}
Now let
\begin{equation}
h(z) = e^{i\gamma(z)}.
\end{equation}
Then~\eref{heq} becomes a difference equation for $\gamma(z)$:
\begin{equation}
\gamma(z - L) = \gamma(z) + \phi(z).
\end{equation}
This equation can be solved straightforwardly via Fourier transform. Alternatively, it determines $\gamma(z)$ recursively on the entire real axis, given an arbitrary choice of $\gamma(z)$ on the interval $0 \le z < L$.

The factor $\prod_{i=1}^N h(z_i)$ is common to all spin components at fixed coordinates. For $|h(z)|=1$, it cancels from coordinate-diagonal probabilities and spin densities. It can, however, affect observables involving spatial coherence, such as the one-particle density matrix away from its diagonal. An explicit construction of $h(z)$ is given in Appendix~\ref{app:few_electron_examples}.

\section{The wavefunction}
\label{wavefunction_sec}

In the previous section, we mapped the nonstationary Kondo model with time-dependent interaction \( J(t) \) to a quantum Knizhnik-Zamolodchikov (qKZ) equation. This mapping allowed us to express the exact time-dependent state of the model in terms of the off-shell Bethe Ansatz solution to the qKZ equation. While this solution is known in the literature, we provide in Appendix~\ref{appendix} a slight but important generalization to the case of  arbitrary ratio $\kappa/c$, where $\kappa$ in the step size in the qKZ equations and $c$ is the crossing parameter $c$ of the XXX $R$-matrix.

We now present the explicit    $N$-particle solution to the nonstationary Schr\"odinger equation for the time-dependent Kondo Hamiltonian with periodic boundary conditions. We focus on the simplest time dependence for which the  model is integrable:
 \begin{equation}
J(t) = \lambda t \pm \sqrt{ \lambda^2 t^2 + \tfrac{16}{3} }.
\label{ourJ}
\en
  Without loss of generality, we set the step parameter in the qKZ equations to \( \kappa = 1 \).

  For this choice and $p(t)=0$, the mapping~\eqref{yg} takes the form
  \begin{equation}
  g(z)=\nu z,\qquad \nu=\frac{1}{cL}=\frac{3\lambda}{8},
  \qquad -c\,g(z)=-\frac{z}{L}.
  \label{eq:linear_mapping}
  \end{equation}

The solution is expressed in terms of the right-moving coordinates 
\beg
z_1=x_1-t, \dots, z_N=x_N-t
\en
 of the conduction electrons.   As described in Section~\re{Sec:N_particle}, the \( N \)-particle state  takes the form
\begin{equation}
\label{npform111}
\ket{\Psi_N} = \prod_{j=1}^{N} \int dx_j\, \hat{\psi}^{\dagger}_{s_j}(x_j)\, \mathcal{A} F_{s_0 \dots s_N}(x_1, \dots, x_N, t) \ket{s_0},
\end{equation}
where \( \mathcal{A} \) denotes antisymmetrization over both electron coordinates \( x_j \) and spin indices \( s_j \), and \( \ket{s_0} \) is the spin state of the impurity. The vector-valued function $F$  with components $F_{s_0 \dots s_N }$  is  constructed from amplitudes defined in each ordering sector of the configuration space.
These amplitudes are labeled by permutations \( Q \) of the impurity and particle labels, and the impurity coordinate is fixed at \( x_0 = 0 \). In each sector, the wavefunction is given by
\begin{equation}
\label{npwf}
F(x_1, \dots, x_N, t) = \sum_Q \theta(x_{Q})\, f^Q_{s_0 \dots s_N}(z_1, \dots, z_N),
\end{equation}
with  
\be
\theta\left(x_Q\right) = 
\begin{cases}
1, & x_{Q(0)} < \dots < x_{Q(N)}, \\
\frac{1}{2},& \text{if $x_{Q(i)}=x_{Q(j)}$ for any pair of $i\ne j$},\\
0, & \text{otherwise}.
\end{cases}
\ee
The amplitudes \( f^Q \) in each sector solve the free Schr\"odinger equation and are obtained from the solution of the qKZ equation, as we now describe.

 The amplitude $f^{N \dots 10} (z_1, \dots, z_N)$ is obtained from the solution of the qKZ equations via \esref{qKz_con_3}, \re{qKZ_con_2}, \re{qKz_con_1}, and \re{offshellstatefinalform}, and reads
\begin{multline}
f^{N \dots 10}(z_1, \dots, z_N)= e^{i\sum_{j=1}^N \gamma(z_j)} \sum_{\{u_j\}}
\prod_{i=0}^N \prod_{j=1}^M \frac{\Gamma\left(-z_i/L - u_j + 1 - ic\right)}{\Gamma\left(-z_i/L - u_j + 1\right)} \\
\times \prod_{1 \leq i < j \leq M} \frac{(u_i - u_j)\, \Gamma(u_i - u_j + ic)}{\Gamma(u_i - u_j - ic + 1)} 
 \prod_{j=1}^M B(\{-z_i/L\}, u_j)  \, \ket{\Omega},
\label{eq:final_wavefunction}
\end{multline}
 where  $z_0\equiv 0$,  and \( \gamma(z) \)   is a phase function defined at the end of Section~\ref{sec:mapping_qkz}.  The phase factor \( e^{i\sum_j \gamma(z_j)} \) is common to all spin components at fixed coordinates and cancels from coordinate-diagonal probabilities, as discussed at the end of Section~\ref{sec:mapping_qkz}.
 
The notation in~\eqref{eq:final_wavefunction} is as follows:
\begin{itemize}

    \item The sum runs over the set  \( \{u_j\}_{j=1}^M \), which are shifted lattice variables: \( u_j = \widetilde{u}_j - l_j \), with \( l_j \in \mathbb{Z} \), and \( \widetilde{u}_j \) arbitrary complex parameters. The     integer \( M \) is  related to the total spin projection by  
    $$
    S^z = \frac{N+1}{2} - M.
   $$
   
    \item $\ket{\Omega}= \ket{\uparrow \dots \uparrow}$ is the fully polarized reference state (vacuum) in which  all spins, including the impurity, point  up.

    \item \( B(\{-z_i/L\}, u_j) \) are  creation operators from the off-diagonal entries of the monodromy matrix, acting on \( \ket{\Omega} \), and defined in Section~\ref{sec_monodromy}.    
    
    \item \( \Gamma(x) \) is the Gamma function, and the crossing parameter \( c \) is an arbitrary real number.
    
\end{itemize}

 The remaining amplitudes \( f^Q \) in other sectors are related to  $f^{N \dots 10} (z_1, \dots, z_N)$ via particle-particle and particle-impurity scattering matrices:
\begin{align}
f^{\dots ji \dots}(z_1,\dots,z_N) & = S^{ij}(z_i, z_j)\, f^{\dots ij \dots}(z_1,\dots,z_N),\quad i,j\ne0,\\
f^{\dots 0j \dots}(z_1,\dots,z_N) & = S^{j0}(z_j)\, f^{\dots j0 \dots}(z_1,\dots,z_N),
\end{align}
where the two-body \( S \)-matrices are given by
\begin{align}
S^{ij}(z_i,z_j) & = \frac{i\nu (z_i - z_j)\, I^{ij} + P^{ij}}{i\nu (z_i - z_j) + 1},\\
S^{j0}(z) &= e^{i\phi(z)}\, \frac{i \nu z\, I^{j0} + P^{j0}}{i \nu z + 1},\\
 e^{i\phi(z)}& = \frac{\pm \sqrt{\nu^2 z^2+\frac{3}{4}} + \frac{i}{2}}{\nu z+i}. \label{explicit_phi}
\end{align}
Here, $I^{ij}$ and $P^{ij}$ are the identity and permutation operators acting on the spins of particles \( i \) and \( j \) (or the impurity), and Eq.~\eqref{explicit_phi} is derived using Eqs.~(\ref{phase}) and (\ref{gJ1})  with \( g(z) = \nu z \).

 \subsection{One-electron example}
\label{sec:example_one_electron}

For one electron, the periodicity condition~\eqref{1pdiffk} can be solved for an arbitrary
real, continuous coupling $J(t)$. To see this explicitly, consider the
electron--impurity singlet and the zero-magnetization triplet,
\begin{equation}
 |s\rangle=\frac{|\uparrow\downarrow\rangle-|\downarrow\uparrow\rangle}{\sqrt{2}},
 \qquad
 |t_0\rangle=\frac{|\uparrow\downarrow\rangle+|\downarrow\uparrow\rangle}{\sqrt{2}},
 \label{eq:example_one_basis}
\end{equation}
where the impurity spin is written first. The scattering matrix~\eqref{smatk} is diagonal
in this basis, with eigenvalues
\begin{equation}
 s_T(z)=e^{i\phi(z)},\qquad
 s_S(z)=e^{i\phi(z)}\frac{ig(z)-1}{ig(z)+1}.
 \label{eq:example_one_eigenvalues}
\end{equation}
Here $g(z)$ and $\phi(z)$ are defined in Eqs.~\eqref{gJ} and~\eqref{phase}. Both eigenvalues have unit modulus for real $J$, with continuous limits understood at $J=0$. Choose unit-modulus functions $h_T$ and
$h_S$ satisfying
\begin{equation}
 h_a(z-L)=s_a(z)h_a(z),\qquad a=T,S.
 \label{eq:example_one_phases}
\end{equation}
Their values on one interval of length $L$ determine them everywhere by
recursion; an explicit choice is given in
Appendix~\ref{app:few_electron_examples}.
Using the sector decomposition~\eqref{1pformk}, an example of a normalized wavefunction on the ring is
\begin{align}
 f^{10}(z)&=\frac{h_T(z)|t_0\rangle+h_S(z)|s\rangle}{\sqrt{2L}},
 \label{eq:example_one_reference}\\
 F(x,t)&=\theta(-x)f^{10}(x-t)
       +\theta(x)S^{10}(x-t)f^{10}(x-t).
 \label{eq:example_one_full}
\end{align}
Equation~\eqref{eq:example_one_phases} gives
$f^{10}(z-L)=S^{10}(z)f^{10}(z)$, which is precisely Eq.~\eqref{1pdiffk}. Moreover, $F^\dagger F=1/L$ away from the contact, so the norm is
one. Although the density is uniform, the relative singlet--triplet phase
changes the amplitudes of the two antiparallel spin configurations. Thus
one-electron scattering can change the spin state. Periodicity determines
how the amplitudes extend around the ring, without imposing a functional
restriction on $J(t)$; there are no compatibility conditions between
different electrons in this sector.

\subsection{Two-electron example}
\label{sec:example_two_electrons}

The two-electron amplitude must satisfy the compatible shift equations~\eqref{2pdiffeqk3} and~\eqref{2pdiffeqk4}. We illustrate their qKZ solution for the positive branch of Eq.~\eqref{ourJ}:
\begin{equation}
 p(t)=0,\qquad
 J(t)=\lambda t+\sqrt{\lambda^2t^2+\frac{16}{3}},\qquad
 c=\frac{8}{3\lambda L}>0,
 \label{eq:example_two_protocol}
\end{equation}
so that $g(z)=z/(cL)$, as in Eq.~\eqref{eq:linear_mapping}. In the sector $x_2<x_1<0$, use the ordered spin basis
\begin{equation}
 e_0=|\downarrow\uparrow\uparrow\rangle,\qquad
 e_1=|\uparrow\downarrow\uparrow\rangle,\qquad
 e_2=|\uparrow\uparrow\downarrow\rangle,
 \label{eq:example_two_basis}
\end{equation}
with spin order (impurity, electron 1, electron 2). One explicit reference
amplitude satisfying both periodicity equations is
\begin{equation}
 f^{210}(z_1,z_2)=\mathcal N h(z_1)h(z_2)
 \left(e^{2\pi iz_1/L}-e^{2\pi iz_2/L}\right)
 \begin{pmatrix}2A+B\\ B-A\\ -A-2B\end{pmatrix},
 \qquad z_j=x_j-t.
 \label{eq:example_two_reference}
\end{equation}
Here $h$ is the scalar factor in Eq.~\eqref{qKZ_con_2}, satisfying Eq.~\eqref{heq} with $|h(z)|=1$, $\mathcal N$ fixes the
normalization, and the scalar functions $A(z_1,z_2;c)$ and $B(z_1,z_2;c)$
are given by the absolutely convergent sums in
Appendix~\ref{app:few_electron_examples}. The remaining sectors follow from Eqs.~\eqref{2psmatk} and~\eqref{eemateqs}. For example,
\begin{equation}
 f^{201}=S^{10}(z_1)f^{210},\qquad
 f^{120}=S^{12}(z_2,z_1)f^{210}.
 \label{eq:example_two_adjacent}
\end{equation}
The three spin components in Eq.~\eqref{eq:example_two_reference} sum to
zero, giving total spin $S=S^z=1/2$. The antisymmetric, separately
$L$-periodic scalar factor makes the reconstructed wavefunction fermionic
without changing its periodicity equations. This is one choice of state:
the coupling alone does not specify the initial wavefunction. In particular,
the node at $x_1=x_2$ is a property of this choice, rather than a requirement
for every spinful two-electron state.

Figure~\ref{fig:example_two_slice} shows a position slice at $t=0$ with
$x_2=-0.4L$ and $c=1$. Writing $F_a$ for the components in
Eq.~\eqref{eq:example_two_basis}, we plot the conditional spatial densities
and spin probabilities
\begin{equation}
 \rho_a(x_1\mid x_2,t)=
 \frac{|F_a(x_1,x_2,t)|^2}
 {\displaystyle\int_{-L/2}^{L/2}dx'_1\sum_{b=0}^2|F_b(x'_1,x_2,t)|^2},
 \qquad
 P_a=\frac{|F_a|^2}{\sum_{b=0}^2|F_b|^2}.
 \label{eq:example_two_probabilities}
\end{equation}
At the impurity, $g(0)=0$ and $S^{10}(0)$ is proportional to the spin
permutation $P^{10}$. Consequently the $e_0$ and $e_1$ probabilities
exchange across $x_1=0$, while their sum and the total density remain
continuous. The conditional impurity polarization is
$\langle S_0^z\rangle_{x_1,x_2}=1/2-P_0$.
The dependence on the driving parameter is explicit through
$c=8/(3\lambda L)$ in the sums and scattering matrices. The scalar phases
$h$ cancel from the probabilities plotted here, although they can affect
observables that probe spatial coherence. The convergent amplitudes are
bounded on the finite ring and define a normalizable state.

\begin{figure}[H]
 \captionsetup{font={color=black},labelfont={color=black}}
 \centering
 \includegraphics[width=0.80\textwidth]{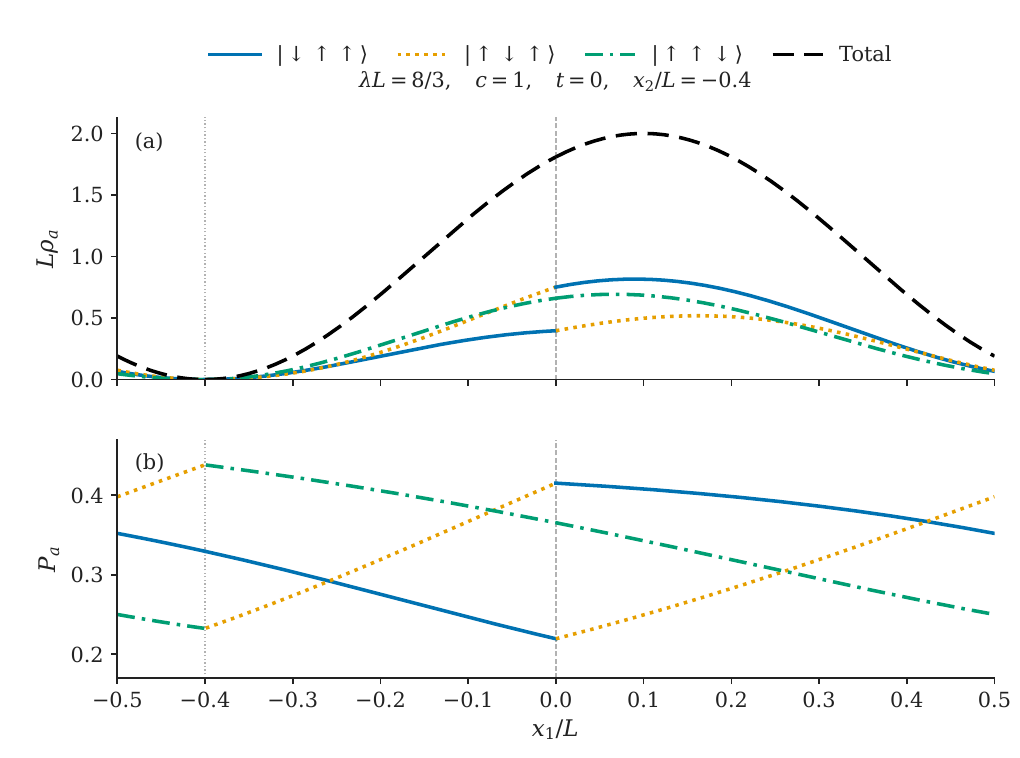}
 \caption{A two-electron wavefunction after imposing spatial periodicity,
 for the state in Eq.~\eqref{eq:example_two_reference} with $p(t)=0$,
 $\lambda L=8/3$ ($c=1$), $t=0$, and $x_2=-0.4L$.
 (a) Spin-resolved conditional densities $L\rho_a$; the black dashed curve is
 their sum. (b) Conditional spin probabilities $P_a$.
 In both panels, blue solid, orange dotted, and green dash-dotted curves
 represent $e_0=|\downarrow\uparrow\uparrow\rangle$,
 $e_1=|\uparrow\downarrow\uparrow\rangle$, and
 $e_2=|\uparrow\uparrow\downarrow\rangle$, respectively.
 Spin order is (impurity, electron 1, electron 2).
 The gray dotted vertical line marks the node at $x_1=x_2$, where $P_a$ is undefined;
 the gray dashed vertical line marks the impurity. Curves on either side of a
 sector boundary are plotted separately. The $e_0$ and $e_1$ components
 exchange across the impurity, while the total density is continuous.
 The wavefunction is spatially periodic, including its complex phase.}
 \label{fig:example_two_slice}
\end{figure}


\section{Discussion}
\label{sec:discussion}

We have obtained here the exact solution of the nonstationary Schr\"odinger equation for the time-dependent Kondo Hamiltonian.
For the model defined with periodic boundary conditions, we showed that the many-body dynamics maps to a system of quantum Knizhnik--Zamolodchikov (qKZ) difference equations for the wavefunction amplitudes, with transport matrices determined by the XXX $R$-matrix and the rational particle--impurity $S$-matrix. We identified a family of time dependences of the spin-exchange coupling for which the evolution is integrable, the simplest being
\beg
J(t) = \lambda t \pm \sqrt{\lambda^2 t^2 + \tfrac{16}{3}}.
\label{j_discussion}
\en
The qKZ equations were then solved exactly using the off-shell Bethe Ansatz, yielding a closed-form expression for the full $N$-particle wavefunction. This provides a complete, explicit description of the real-time evolution in a nontrivial quantum impurity model far from 
equilibrium.

  Several interesting physical scenarios can be explored using our solution. As an illustration, consider the case $\lambda>0$ with the plus sign in \eref{j_discussion}, and let the system evolve from a large negative time $t=-T$ to a large positive time $t=+T$. This protocol corresponds to gradually switching on the antiferromagnetic Kondo interaction: at early times the coupling behaves as $8/(3\lambda |t|)$, while at late times it grows linearly as $2\lambda t$. Suppose the impurity spin is initially polarized up, $S^z_\text{imp}(t=-T)=+\tfrac{1}{2}$, and the conduction electrons are prepared in their Fermi liquid ground state. Our exact solution provides access to the full many-body wavefunction at $t=+T$, from which one can evaluate observables such as the impurity polarization $\langle S^z_\text{imp}\rangle$, the spatial profile of the electron polarization $\langle S^z_\text{el}(x)\rangle$, and the correlation function $\langle S^z_\text{el}(x)\,S^z_\text{imp}\rangle$ directly related to the Knight shift in NMR experiments. More global quantities, including the von Neumann entanglement entropy ${\cal S}_\text{ent}$ between the impurity and the conduction band and the electron momentum distribution function, can also be computed.  

While the exact wavefunction has a complicated structure, past experience indicates that substantial simplifications occur at asymptotically large times. In the BCS model with a coupling inversely proportional to time, for instance, it was possible to obtain compact expressions for the full wavefunction and all local observables in terms of elementary functions at $t\to+\infty$~\cite{Zabalo2022}. Although the present Kondo problem is technically more intricate, owing to the richer algebraic structure of its solution, we likewise expect that explicit and reasonably simple results can be obtained for the observables of interest in the large-time regime. Two limiting cases already illustrate the possibilities. In the adiabatic limit $\lambda\to 0$, the system remains in the instantaneous ground state, giving $\langle S^z_\text{imp}\rangle=0$ and ${\cal S}_\text{ent}=\ln 2$, reflecting the formation of a singlet between the impurity and an electron at $x=0$ in the $J\to+\infty$ ground state. In the opposite, diabatic limit $\lambda\to+\infty$, the system remains frozen in its initial configuration, so that $\langle S^z_\text{imp}\rangle=\tfrac{1}{2}$ and ${\cal S}_\text{ent}=0$ at $t=+T$. Between these limits, we anticipate a rich dependence of observables on the adiabaticity parameter $\lambda$, with the possibility of a sharp transitions at a critical value. This line of inquiry naturally links to experimentally relevant probes of non-equilibrium Kondo physics, ranging from NMR experiments to quantum-dot systems and ultracold-atom realizations~\cite{Grobis2007,Bauer2013,Amaricci2025}.

We expect the methods developed in this work to apply equally well to time-dependent extensions of other one-dimensional integrable quantum field theories~\cite{EsslerKonik2005,ThackerRMP1981}. Our construction of the general solution of the nonstationary Schr\"odinger equation relies only on two key ingredients: linear dispersion and the integrability  of the model. Under suitable constraints on the time dependence of the interaction, the boundary conditions are expected to give rise to integrable matrix difference equations  analogous to the qKZ equations analyzed here. Natural candidates for such extensions include the Gross--Neveu \cite{Gross1974,Zamolodchikov1979,Andrei1979,Destri1982} and Thirring \cite{Thirring1958,Coleman1975,Bergknoff1979,Korepin1979, ThackerRMP1981} models, both of which are richer and more intricate than the Kondo model: the Gross--Neveu model features fermionic bulk interactions, a running coupling constant and a spectrum of bound states, while the Thirring model is built on current--current interactions and possesses a distinct internal symmetry structure.

\section*{Acknowledgments}
We thank Natan Andrei for helpful discussions.
We used ChatGPT (OpenAI) to revise and polish the language of the manuscript. All scientific content was developed by the authors.

 
 \begin{appendix}
\numberwithin{equation}{section}

\section{Off-shell Bethe Ansatz and solution to qKZ equations}
\label{appendix}

In the main text, we showed that for certain time dependences of the interaction strength $J(t)$, the periodic boundary conditions imposed on the many-body wavefunction of the time-dependent Kondo Hamiltonian reduce to a set of quantum Knizhnik--Zamolodchikov (qKZ) equations:
\be
\varphi(y_0,\dots, y_j + \kappa,\dots, y_N) 
= M_j(y_0,\dots,y_N)\, 
\varphi(y_0,\dots,y_N),
\label{qKZap}
\ee
where the transport operator $M_j(y_0,\dots,y_N)$ is defined as
\begin{align}
M_j(y_0,\dots,y_N) &= 
R^{j+1\,j}(y_{j+1} - y_j - \kappa)\, R^{j+2\,j}(y_{j+2} - y_j - \kappa) \cdots R^{N\,j}(y_N - y_j - \kappa)
\nonumber\\
&\quad \times R^{0\,j}(y_0 - y_j)\, R^{1\,j}(y_1 - y_j) \cdots R^{j-1\,j}(y_{j-1} - y_j).
\label{transportopap}
\end{align}
In this appendix, we construct an explicit solution to the qKZ equations using the off-shell Bethe Ansatz method. This approach was originally developed by Babujian~\cite{Babujian_1997} in the special case where the step and crossing parameter are fixed to $\kappa = 2$ and $c = 1$, respectively. 

However, since the time-dependent interaction $J(t)$ in our model depends only on the ratio $\kappa/c$, it is desirable to keep this ratio arbitrary to accommodate a wider class of solvable couplings. We therefore generalize Babujian's method to allow for an arbitrary $\kappa/c$ ratio. Specifically, without loss of generality, we fix the step to $\kappa = 1$ and treat $c$ as a free parameter throughout the construction.

\subsection{Monodromy matrix}
\label{sec_monodromy}

We begin by recalling the form of the XXX $R$-matrix introduced earlier in Eq.~\re{rmatdef}. It can be written as
\begin{equation}
R_{ab}(x) = b(x)\, I_{ab} + c(x)\, P_{ab}, \qquad
b(x) = \frac{ix}{ix + c}, \qquad
c(x) = \frac{c}{ix + c},
\label{rmatap}
\end{equation}
where $c$ is the crossing parameter, $I_{ab}$ is the identity operator, and $P_{ab}$ is the permutation operator acting on the tensor product of spin spaces $a$ and $b$.

This $R$-matrix satisfies the following properties:
\begin{itemize}
  \item \textbf{Initial condition:} $R_{ab}(0) = P_{ab}$.
  \item \textbf{Unitarity:} $R_{ab}(x) R_{ab}(-x) = \mathbb{1}$.
  \item \textbf{Crossing symmetry:}
  \begin{equation}
  R_{ab}(x) = -\sigma^y_a R^{t_a}_{ab}(-x + ic)\, \sigma^y_a \left( \frac{x}{-x + ic} \right),
  \end{equation}
\end{itemize}
where $t_a$ denotes transposition in space $a$ and $\sigma^y_a$ is the Pauli matrix acting in that space.

The first step in the off-shell Bethe Ansatz procedure is to define the \textit{monodromy matrix},
\begin{equation}
\label{monodromy}
T_{N \dots 10\, a}(\{y_i\}, y) = R_{0a}(y_0 - y)\, R_{1a}(y_1 - y)\cdots R_{Na}(y_N - y),
\end{equation}
which acts on the tensor product space
\[
V_0 \otimes V_1 \otimes \cdots \otimes V_N \otimes V_a,
\]
where $V_1, \dots, V_N$ correspond to the $N$ particles, $V_0$ corresponds to the impurity, and $V_a$ is the auxiliary space. Since we are working with spin-$\frac{1}{2}$ particles, each space $V_\alpha \cong \mathbb{C}^2$ for $\alpha = a, 0, 1, \dots, N$. Here $\{y_i\} \equiv (y_0,y_1, \dots, y_j, \dots, y_N)$ represents the set of anisotropy parameters which are related to $z_j$ through the relations (\ref{yg}). The operators $R_{ia}(x)$ are the XXX $R$-matrices defined in Eq.~\re{rmatap}.

The $R$-matrices satisfy the Yang--Baxter algebra:
\begin{equation}
\begin{split}
R^{ij}(y_i - y_j)\, R^{ik}(y_i - y_k)\, R^{jk}(y_j - y_k)
= R^{jk}(y_j - y_k)\, R^{ik}(y_i - y_k)\, R^{ij}(y_i - y_j).
\end{split}
\label{YB4}
\end{equation}
 Using these relations, one can show that the monodromy matrices defined in Eq.~\re{monodromy} also satisfy the Yang--Baxter algebra:
\begin{equation}
\begin{split}
&T_{N \dots 10\, a}(\{y_i\}, y_a)\, T_{N \dots 10\, b}(\{y_i\}, y_b)\, R_{ab}(y_a - y_b) \\
&\quad = R_{ab}(y_a - y_b)\, T_{N \dots 10\, b}(\{y_i\}, y_b)\, T_{N \dots 10\, a}(\{y_i\}, y_a),
\end{split}
\label{YBM}
\end{equation}
where $a$ and $b$ are auxiliary spaces. This algebraic structure ensures the mutual commutativity of transfer matrices and is a cornerstone of the integrability framework.

Let the monodromy matrix take the following form in the auxiliary space:
\begin{equation}
T_{N \dots 10\, a}(\{y_i\}, y) =
\begin{pmatrix}
A(\{y_i\}, y) & B(\{y_i\}, y) \\
C(\{y_i\}, y) & D(\{y_i\}, y)
\end{pmatrix},
\label{formM}
\end{equation}
where the operators $X(\{y_i\}, y)$, with $X = A, B, C, D$, act in the tensor product space $V_0 \otimes V_1 \otimes \cdots \otimes V_N$ corresponding to the impurity ($0$) and particles ($1,\dots,N$). The $R$-matrix $R_{ia}(x)$ is defined in Eq.~\re{rmatap}.
Using the decomposition~\re{formM} in the Yang--Baxter relation~\re{YBM}, one obtains the following commutation relations among the operators $A, B, C, D$ (we suppress the arguments $\{y_i\}$ for readability):
\begin{align}\label{commrel1ap}
A(y_a)\, A(y_b) &= A(y_b)\, A(y_a), \\
B(y_a)\, B(y_b) &= B(y_b)\, B(y_a), \\
C(y_a)\, C(y_b) &= C(y_b)\, C(y_a), \\
D(y_a)\, D(y_b) &= D(y_b)\, D(y_a),
\end{align}
 \begin{align}
A(y_a)\, B(y_b) &= \frac{1}{b(y_a - y_b)}\, B(y_b)\, A(y_a)
- \frac{c(y_a-y_b)}{b(y_a - y_b)}\, B(y_a)\, A(y_b), \\
D(y_a)\, B(y_b) &= \frac{1}{b(y_b - y_a)}\, B(y_b)\, D(y_a)
- \frac{c(y_b - y_a)}{b(y_b - y_a)}\, B(y_a)\, D(y_b),\label{commrel2ap}
\end{align}

where the scalar functions $b(x)$ and $c(x)$ are defined in Eq.~\re{rmatap}.

 Consider a reference state $\ket{\Omega}$ in which all particle and impurity spins point along the positive $z$-axis:
\begin{equation}
\ket{\Omega} = \ket{\uparrow \dots \uparrow}.
\end{equation}
This is an eigenvector of the operators $A(\{y_i\}, y)$ and $D(\{y_i\}, y)$ with
\begin{align}
A(\{y_i\}, y)\, \ket{\Omega} &= \ket{\Omega}, \label{eigenvalue1} \\
D(\{y_i\}, y)\, \ket{\Omega} &= \prod_{i=0}^{N} b(y_i - y)\, \ket{\Omega}. \label{eigenvalue2}
\end{align}
 The operator $C(\{y_i\}, y)$ acts trivially on the reference state:
\begin{equation}
C(\{y_i\}, y)\, \ket{\Omega} = 0,
\end{equation}
while the operator $B(\{y_i\}, y)$ acts nontrivially, flipping one spin:
\begin{equation}
B(\{y_i\}, y)\, \ket{\Omega} = \sum_{i=0}^{N} c(y_i - y)\, \prod_{k=i+1}^{N}b(y_k-y)\, \sigma_i^- \ket{\Omega}.
\label{eq:B_on_vacuum}
\end{equation}

By acting on $\ket{\Omega}$ with $M$ spin-flipping operators $B(\{y_i\}, u_\alpha)$, where $\alpha = 1, \dots, M$, one constructs a state with $M$ down spins:
\begin{equation}
\ket{\Omega'} = \prod_{\alpha=1}^{M} B(\{y_i\}, u_\alpha)\, \ket{\Omega}.
\label{onshellstate}
\end{equation}
This state has total $z$-component of spin
\begin{equation}
S^z = \frac{N+1}{2} - M.
\end{equation}
In the standard (on-shell) Bethe Ansatz, the parameters $u_\alpha$ are called \textit{Bethe roots}. These must all be distinct in order for the state $\ket{\Omega'}$ to be nonzero.

   \subsection{Step monodromy matrix}

The second step in this procedure is to define the \textit{step monodromy matrix}
\beg
\begin{split}
T^{S_j}_{N\dots 10\,a}(\{y_i\}, y) &= R_{0a}(y_0 - y)\, R_{1a}(y_1 - y) \dots R_{j-1\,a}(y_{j-1} - y) \\
&\quad \times R_{j\,a}(y_j - y)\, R_{j+1\,a}(y_{j+1} - y - 1) \dots R_{N\,a}(y_N - y - 1),
\end{split}
\label{stepmonodromy}
\en
where we set the step $\kappa = 1$. The superscript $S_j$ denotes a ``step at site $j$'', indicating that the spectral parameter is shifted by one unit after the $j$-th term in the product.   Although the step monodromy matrix~\eqref{stepmonodromy} is, in general, different from the usual monodromy matrix~\eqref{monodromy}, for $j = N$ they coincide:
\begin{equation}
T^{S_N}_{N\dots 10\,a}(\{y_i\}, y_N) = T_{N\dots 1\,a}(\{y_i\}, y_N).
\label{equivtmats}
\end{equation}

The trace of the step monodromy matrix~\eqref{stepYB} in the auxiliary space gives the operator $M_j$ defined in~\eref{transportopap}:
\begin{equation}
M_j = \mathrm{tr}_a\, T^{S_j}_{N\dots 10\,a}(\{y_i\}, y_j).
\label{relzt}
\end{equation}
Unlike the standard Yang-Baxter relation satisfied by monodromy matrices~\eqref{YBM}, the step monodromy matrices satisfy a modified Yang-Baxter equation:
\beg
\begin{split}
&T^{S_j}_{N\dots 10\,a}(\{y_i\}, y_j)\, T_{N\dots 10\,b}(\{y_i\}, y)\, R_{ab}(y_j+1 - y) \\
&= R_{ab}(y_j - y)\, T_{N\dots 10\,b}(\{y_i\}_j, y)\, T^{S_j}_{N\dots 10\,a}(\{y_i\}, y_j),
\end{split}
\label{stepYB}
\en 
where 
\beg
\{y_i\}_j \equiv (y_0,\dots, y_j + 1, \dots, y_N)
\en
 denotes the set of anisotropy parameters with $y_j$ shifted by one unit. The step monodromy matrix takes the following form in the auxiliary space:
\begin{equation}
T^{S_j}_{N\dots 10\,a}(\{y_i\}, y) =
\begin{pmatrix}
A^{S_j}(\{y_i\}, y) & B^{S_j}(\{y_i\}, y) \\
C^{S_j}(\{y_i\}, y) & D^{S_j}(\{y_i\}, y)
\end{pmatrix}.
\label{formSM}
\end{equation}
Using this and \eref{formM} in the Yang-Baxter equation~\eqref{stepYB}, we obtain the following commutation relations among the operators $X^{S_j}(y)$ with $X = A, B, C, D$. (We suppress the dependence on $\{y_i\}$ for readability and introduce the notation $X_j(y)\equiv X(\{y\}_j,y)$, $X=A,B,C,D$.)
\begin{align}\label{commrel3ap}
A^{S_j}(y_j) A(y) &= A_j(y) A^{S_j}(y_j), \\
C^{S_j}(y_j) C(y) &= C_j(y) C^{S_j}(y_j), \\
B^{S_j}(y_j) B(y) &= B_j(y) B^{S_j}(y_j), \\
D^{S_j}(y_j) D(y) &= D_j(y) D^{S_j}(y_j), \\
A^{S_j}(y_j) B(y) &= \frac{1}{b(y_j - y + 1)} B^{S_j}(y_j) A(y)
+ \frac{1}{c(y_j - y + 1)} A_j(y) B^{S_j}(y_j), \\
A^{S_j}(y_j) B(y) &= \frac{1}{b(y_j - y + 1)} B_j(y) A^{S_j}(y_j)
- \frac{c(y_j - y + 1)}{b(y_j - y + 1)} B^{S_j}(y_j) A(y), \\
C^{S_j}(y_j) A(y) &= b(y_j - y) A_j(y) C^{S_j}(y_j)
+ c(y_j - y) C_j(y) A^{S_j}(y_j), \\
A^{S_j}(y_j) C(y) &= b(y_j - y) C_j(y) A^{S_j}(y_j)
+ c(y_j - y) A_j(y) C^{S_j}(y_j), \\
D^{S_j}(y_j) B(y) &= b(y - y_j) B_j(y) D^{Sj}(y_j)
- \frac{c(y - y_j)}{b(y - y_j)} B^{S_j}(y_j) D(y).
\label{commrel4ap}
\end{align}

\subsection{Ansatz for the amplitude}

The third and final step in the procedure is to choose an Ansatz for the amplitude $\varphi(y_0,\dots,y_N)$ that provides a solution to the qKZ equation~(\ref{qKZap}). 
The Ansatz takes the following form:
\beg
\varphi(y_0,\dots, y_N)=\sum_{u_j}\prod_{j=1}^{M}B(\{y_i\},u_j)\,w(\{y_i\},\{u_j\})\,\ket{\Omega}.
\label{offshellstate}
\en
The total $z$-component of the spin of this state is $S^z = \frac{N+1}{2} - M$. 
Here, $w(\{y_i\},\{u_j\})$ is an unknown function that needs to be determined. 
Comparing the general state in the traditional on-shell Bethe Ansatz approach~(\ref{onshellstate}) with the above off-shell state~(\ref{offshellstate}), we see that the parameters $u_j$, $j=1,\dots,M$, are associated with the Bethe roots. 
Unlike in the on-shell Bethe Ansatz, here these parameters are not constrained by Bethe equations but are instead summed over:
\be 
u_j = \widetilde{u}_j - l_j, \quad l_j \in \mathbb{Z}.
\label{summation}
\ee
The summation is over the integers $l_j$, while the parameters $\widetilde{u}_j$, $j = 1, \dots, M$, are arbitrary constants.

We now turn to the right-hand side of the qKZ equations~(\ref{qKZap}).  
From equation~(\ref{relzt}), the operator $M_j$ can be expressed as the trace of the step monodromy matrix over the auxiliary space. Using its explicit form~(\ref{formSM}), we find
\be
M_j = \text{tr}_a\, T^{S_j}_{N\dots10\,a}(\{y_i\}, y_j) = A^{S_j}(\{y_i\}, y_j) + D^{S_j}(\{y_i\}, y_j).
\label{traceform}
\ee

Due to the periodicity condition~(\ref{periodic1}), it suffices to consider only the transport operator $Z_N$, which corresponds to the action of $M_N$ in the qKZ equations. Substituting the Ansatz for the amplitude $\varphi(y_0, \dots, y_N)$ from equation~(\ref{offshellstate}) and the trace form~(\ref{traceform}) for the operator $M_N$ into right-hand side of the qKZ equations~(\ref{qKZap}), we proceed by applying the standard off-shell Bethe Ansatz technique~\cite{Babujian_1997,SklyaninQISM}.
This method involves commuting the operators $A^{S_N}(\{y_i\}, y_N)$ and $D^{S_N}(\{y_i\}, y_N)$ past the $B(\{y_i\}_N, u_j)$ operators in the amplitude. In doing so, we repeatedly use the commutation relations~(\ref{commrel1ap})--(\ref{commrel4ap}). For a detailed account, we refer the reader to the excellent textbook on Bethe Ansatz~\cite{ODBA} and to the original works~\cite{Babujian_1997,SklyaninQISM}.  

Carrying out this procedure, we obtain the following expression:
\begin{align}
&\text{tr}_a\, T^{S_N}_{N\dots10\,a}(\{y_i\}, y_N)\, \varphi(y_0, \dots, y_N)
\notag \\
&= \left(A^{S_N}(\{y_i\}, y_N) + D^{S_N}(\{y_i\}, y_N)\right) \varphi(y_0, \dots, y_N)
\notag \\
&= \sum_{u_j} \prod_{j=1}^M B(\{y_i\}_N, u_j)\, w(\{y_i\}, \{u_j\}) \left[ \prod_{k=1}^M \frac{1}{b(y_N + 2 - u_k)} A^{S_N}(\{y_i\}, y_N) \right. \notag \\
&\quad\left. + \prod_{k=1}^M \frac{1}{b(u_k - y_N)} D^{S_N}(\{y_i\}, y_N) \right] \ket{\Omega} + \sum_{u_j} \sum_{m=1}^M \left(UW_A^m + UW_D^m\right),
\label{offshellstateaction}
\end{align}
where the last two terms, $UW_A^m$ and $UW_D^m$, are known as the unwanted terms and are given explicitly by
\begin{multline}
UW_A^m = -\frac{c(y_N + 1 - u_m)}{b(y_N + 1 - u_m)}\, B^{S_N}(\{y_i\}, y_m) \prod_{\substack{j=1 \\ j\ne m}}^M B(\{y_i\}, u_j) \\
\times   \prod_{k\neq m,k=1}^{M}\frac{1}{b(u_m - u_k)}\, A(\{y_i\}, u_m)\, w(\{y_i\}, \{u_j\}) \ket{\Omega},
\end{multline}
\begin{multline}
UW_D^m = -\frac{c(u_m - y_N)}{b(u_m - y_N)}\, B^{S_N}(\{y_i\}, y_m) \prod_{\substack{j=1 \\ j\ne m}}^M B(\{y_i\}, u_j)\times \\
\times   \prod_{k\neq m,k=1}^{M}\frac{1}{b(u_k - u_m)}\, D(\{y_i\}, u_m)\, w(\{y_i\}, \{u_j\}) \ket{\Omega}.
\end{multline}

As seen in equation~(\ref{offshellstateaction}), the action of $M_N$ on the amplitude yields four distinct terms. The second term vanishes by virtue of the identity $D^S(\{y_i\}, y_N)\ket{\Omega} = 0$, which follows from equations~(\ref{eigenvalue2}), (\ref{equivtmats}), and~(\ref{relzt}). The first term has the same operator structure as the off-shell state~\re{offshellstate} itself, involving a product over $B$ operators. The remaining two terms, $UW_A^m$ and $UW_D^m$, must be canceled to obtain a valid solution.

This cancellation can be achieved by appropriately choosing the function $w(\{y_i\}, \{u_j\})$ such that the unwanted terms cancel pairwise. We propose the following Ansatz:
\be
w(\{y_i\}, \{u_j\}) = \prod_{i=0}^N \prod_{j=1}^M \eta(y_i - u_j)\!\!\! \prod_{1 \le i < j \le M} \tau(u_i - u_j),
\label{defX}
\ee
where the functions $\eta(x)$ and $\tau(x)$ satisfy
\be
b(x)\, \eta(x) = \eta(x - 1), \qquad \frac{\tau(x)}{b(x)} = \frac{\tau(x - 1)}{b(1 - x)}.
\label{constraintspsitau}
\ee

One can verify that with this choice,
\be
UW_A^m(\{y_i\}, \{u_j\}_m)\, w(\{y_i\}, \{u_j\}_m) = -UW_D^m(\{y_i\}, \{u_j\})\, w(\{y_i\}, \{u_j\}),
\label{unwantedcancel}
\ee
where $\{u_j\}_m$ denotes the set with $u_m$ replaced by $u_m + 1$. Summing over the discrete shifts defined in equation~(\ref{summation}) then ensures that the unwanted terms cancel, leaving only the first term on the right-hand side of equation~(\ref{offshellstateaction}).

We are thus left with the simplified expression
\be
\text{tr}_a\, T^{S_N}_{N\dots10\,a}(\{y_i\}, y_N)\, \varphi(y_0, \dots, y_N) = \sum_{u_j} \prod_{j=1}^M B(\{y_i\}_N, u_j)\, w(\{y_i\}, \{u_j\}) \prod_{k=1}^M \frac{1}{b(y_N + 1 - u_k)} \ket{\Omega}.
\label{solution1kz}
\ee
Using the functional relation~(\ref{constraintspsitau}), we confirm that the right-hand side of equation~(\ref{solution1kz}) matches the left-hand side of the qKZ equation~(\ref{qKZap}). Therefore,
\be
M_N\, \varphi(y_0, \dots, y_N) = \varphi(y_0, \dots, y_N + 1),
\ee
as required by~(\ref{relzt}). This establishes that the Ansatz~(\ref{offshellstate}) solves the qKZ equation~(\ref{qKZap}), provided the functions $\eta(x)$ and $\tau(x)$ satisfy the constraints~(\ref{constraintspsitau}).

Solving these functional equations using the definition of $b(x)$ from~(\ref{rmatap}), we find the explicit solutions
\be
\eta(x) = \frac{\Gamma(x + 1 - ic)}{\Gamma(x + 1)}, \qquad \tau(x) = x\, \frac{\Gamma(x + ic)}{\Gamma(x - ic + 1)}.
\label{solutionpsitau}
\ee

Substituting into~(\ref{defX}), we obtain the final form of the amplitude:
\begin{multline}
\varphi(y_0, \dots, y_N) = \sum_{u_j} \prod_{j=1}^M B_{N\dots10}(\{y_i\}, u_j) \prod_{i=0}^N \prod_{j=1}^M \frac{\Gamma(y_i - u_j + 1 - ic)}{\Gamma(y_i - u_j + 1)} \\
\times \prod_{1 \le i < j \le M} \frac{(u_i - u_j)\, \Gamma(u_i - u_j + ic)}{\Gamma(u_i - u_j - ic + 1)} \ket{\Omega}.
\label{offshellstatefinalform}
\end{multline}


\section{Details of the one- and two-electron examples}
\label{app:few_electron_examples}

\subsection{Scalar phases}

The scalar difference equation~\eqref{eq:example_one_phases}, which generalizes Eq.~\eqref{heq} to either spin channel, can be solved by specifying a seed on one interval.
For example, choose a real $\delta_a(L)$ such that
$e^{i\delta_a(L)}=s_a(L)$ and set
\begin{equation}
 h_a(w)=e^{-iw\delta_a(L)/L},\qquad 0\leq w<L.
 \label{eq:example_phase_seed}
\end{equation}
For any positive integer $m$, extend this seed by
\begin{align}
 h_a(w+mL)&=h_a(w)\prod_{r=1}^{m}s_a(w+rL)^{-1},\\
 h_a(w-mL)&=h_a(w)\prod_{r=0}^{m-1}s_a(w-rL).
 \label{eq:example_phase_products}
\end{align}
These finite products give an explicit solution for every real $z$ and
every real continuous protocol $J(t)$. The seed makes the extension
continuous at the interval endpoints, and $|h_a(z)|=1$. For the two-electron
example use $s_a=e^{i\phi}$, namely the triplet factor. On the positive
branch in Eq.~\eqref{eq:example_two_protocol}, one may take
$\phi(z)=-2\arctan[J(-z)/4]$, consistently with Eq.~\eqref{phase}.
Other compatible seeds describe different states.

\subsection{Explicit two-electron amplitude}

Define
\begin{align}
 K_n(z_1,z_2;c)
 &=\frac{\Gamma(n-ic)}{\Gamma(n+1)}
   \prod_{j=1}^{2}
   \frac{\Gamma(n-z_j/L-ic)}{\Gamma(n-z_j/L+1)},\qquad n\geq0,
 \label{eq:example_kernel}\\
 I_0&=\sum_{n=0}^{\infty}K_n,\qquad
 I_1=\sum_{n=0}^{\infty}nK_n.
 \label{eq:example_moments}
\end{align}
The functions entering Eq.~\eqref{eq:example_two_reference} are
\begin{align}
 A&=\left(-\frac{z_1}{L}+ic\right)
       \left(I_1-\frac{z_2}{L}I_0\right),\\
 B&=\left(\frac{z_1-z_2}{L}+ic\right)(I_1-icI_0).
 \label{eq:example_ab}
\end{align}
These formulas use the qKZ arguments $-cg(z_j)=-z_j/L$ from Eqs.~\eqref{yg} and~\eqref{eq:linear_mapping}, with
$z_0=0$ and unit qKZ step.

To obtain them from Eq.~\eqref{eq:final_wavefunction}, equivalently Eq.~\eqref{offshellstatefinalform}, set $N=2$ and $M=1$ and choose the summation lattice $u=-n$ in Eq.~\eqref{summation}. Its scalar weight is
\begin{equation}
 W_n=\prod_{j=0}^{2}
 \frac{\Gamma(n-z_j/L+1-ic)}{\Gamma(n-z_j/L+1)}.
\end{equation}
For the monodromy ordering in Eq.~\eqref{monodromy}, introduce
\begin{equation}
 b_j=\frac{n-z_j/L}{n-z_j/L-ic},\qquad
 d_j=\frac{-ic}{n-z_j/L-ic}.
\end{equation}
By Eq.~\eqref{eq:B_on_vacuum}, the three components of $B\lvert\Omega\rangle$ are
$(d_0b_1b_2,d_1b_2,d_2)^{\mathsf T}$. Using
$\Gamma(v+1)=v\Gamma(v)$ gives
\begin{equation}
 W_n B\lvert\Omega\rangle=-icK_n
 \begin{pmatrix}
 (n-z_1/L)(n-z_2/L)\\
 (n-ic)(n-z_2/L)\\
 (n-ic)(n-z_1/L-ic)
 \end{pmatrix}.
 \label{eq:example_unprojected}
\end{equation}
Negative $n$ terms vanish because $1/\Gamma(n+1)=0$.
Project onto total spin $1/2$ with
$\Pi_{1/2}=I-\mathbf e\mathbf e^{\mathsf T}/3$, where
$\mathbf e=(1,1,1)^{\mathsf T}$ is the spin-$3/2$ vector in this
magnetization sector. The projection removes the common quadratic term
in $n$ in Eq.~\eqref{eq:example_unprojected}. Summation then gives
$(2A+B,B-A,-A-2B)^{\mathsf T}$, up to a constant absorbed into
$\mathcal N$. Since the scattering matrices are $SU(2)$ invariant, this
projection preserves the qKZ equations. One may first carry out the
construction for $\operatorname{Im}c<0$, where the unprojected sums
converge, and then analytically continue the projected expression to
real $c>0$.

Indeed, the Gamma-ratio asymptotic formula
\begin{equation}
 \frac{\Gamma(n+\alpha)}{\Gamma(n+\beta)}
 =n^{\alpha-\beta}\bigl[1+O(n^{-1})\bigr]
\end{equation}
implies
\begin{equation}
 K_n=n^{-3-3ic}\bigl[1+O(n^{-1})\bigr].
\end{equation}
For real $c$, $|n^{-3ic}|=1$. Thus $|K_n|=O(n^{-3})$ and
$|nK_n|=O(n^{-2})$, proving absolute convergence of both sums. The
convergence is uniform on compact sets of real coordinates. Ratios
involving a pole of the denominator Gamma function are understood
using the entire reciprocal Gamma function.

Writing $v=(2A+B,B-A,-A-2B)^{\mathsf T}$, the resulting amplitude satisfies both spatial periodicity equations~\eqref{2pdiffeqk3} and~\eqref{2pdiffeqk4}. It also obeys
$v(z_2,z_1)=P^{12}S^{12}(z_2,z_1)v(z_1,z_2)$.
Together with the antisymmetric scalar factor in
Eq.~\eqref{eq:example_two_reference}, this ensures
$F(x_2,x_1,t)=-P^{12}F(x_1,x_2,t)$ after reconstructing the sectors.
For the plotted slice, only sectors $120$, $210$, and $201$ are needed,
so Eq.~\eqref{eq:example_two_adjacent} suffices.

As a numerical check, at $c=1$, $t=0$, $x_1=-0.2L$, and $x_2=-0.4L$, the
three conditional spin probabilities are
\begin{equation}
 (P_0,P_1,P_2)=(0.275385,\,0.318956,\,0.405659).
\end{equation}
These numbers follow from the three components of $v$; the common factors
$\mathcal N$, $h(z_1)h(z_2)$, and
$e^{2\pi iz_1/L}-e^{2\pi iz_2/L}$ cancel from their ratios.
At the same coordinates, repeating the construction with $c=1/2$
($\lambda L=16/3$) gives $(P_0,P_1,P_2)=(0.347061,0.266911,0.386028)$.
This compares selected states for two protocols; a common initial state
has not been imposed.


\end{appendix}


\begin{thebibliography}{132}

\bibitem{GeorgescuRMP2014}
I. M. Georgescu, S. Ashhab and F. Nori, \emph{Quantum simulation}, Rev. Mod. Phys. \textbf{86}, 153 (2014),
doi:\href{https://doi.org/10.1103/RevModPhys.86.153}{10.1103/RevModPhys.86.153} [preprint
doi:\href{https://doi.org/10.48550/arXiv.1308.6253}{10.48550/arXiv.1308.6253}].

\bibitem{GrossBlochScience2017}
C. Gross and I. Bloch, \emph{Quantum simulations with ultracold atoms in optical lattices}, Science
\textbf{357}, 995 (2017), doi:\href{https://doi.org/10.1126/science.aal3837}{10.1126/science.aal3837}.

\bibitem{BlattRoosNatPhys2012}
R. Blatt and C. F. Roos, \emph{Quantum simulations with trapped ions}, Nat. Phys. \textbf{8}, 277 (2012),
doi:\href{https://doi.org/10.1038/nphys2252}{10.1038/nphys2252}.

\bibitem{HouckTureciKochNatPhys2012}
A. A. Houck, H. E. Türeci and J. Koch, \emph{On-chip quantum simulation with superconducting circuits}, Nat.
Phys. \textbf{8}, 292 (2012), doi:\href{https://doi.org/10.1038/nphys2251}{10.1038/nphys2251}.

\bibitem{MonroeRMP2021}
C. Monroe et al., \emph{Programmable quantum simulations of spin systems with trapped ions}, Rev. Mod. Phys.
\textbf{93}, 025001 (2021),
doi:\href{https://doi.org/10.1103/RevModPhys.93.025001}{10.1103/RevModPhys.93.025001} [preprint
doi:\href{https://doi.org/10.48550/arXiv.1912.07845}{10.48550/arXiv.1912.07845}].

\bibitem{GongScience2021}
M. Gong et al., \emph{Quantum walks on a programmable two-dimensional 62-qubit superconducting processor},
Science \textbf{372}, 948 (2021), doi:\href{https://doi.org/10.1126/science.abg7812}{10.1126/science.abg7812}
[preprint doi:\href{https://doi.org/10.48550/arXiv.2102.02573}{10.48550/arXiv.2102.02573}].

\bibitem{SatzingerScience2021}
K. J. Satzinger et al., \emph{Realizing topologically ordered states on a quantum processor}, Science
\textbf{374}, 1237 (2021), doi:\href{https://doi.org/10.1126/science.abi8378}{10.1126/science.abi8378}
[preprint doi:\href{https://doi.org/10.48550/arXiv.2104.01180}{10.48550/arXiv.2104.01180}].

\bibitem{CongPRX2022}
I. Cong, H. Levine, A. Keesling, D. Bluvstein, S.-T. Wang and M. D. Lukin, \emph{Hardware-efficient,
fault-tolerant quantum computation with Rydberg atoms}, Phys. Rev. X \textbf{12}, 021049 (2022),
doi:\href{https://doi.org/10.1103/PhysRevX.12.021049}{10.1103/PhysRevX.12.021049}.

\bibitem{MiScience2024}
X. Mi et al., \emph{Stable quantum-correlated many-body states through engineered dissipation}, Science
\textbf{383}, 1332 (2024), doi:\href{https://doi.org/10.1126/science.adh9932}{10.1126/science.adh9932}
[preprint doi:\href{https://doi.org/10.48550/arXiv.2304.13878}{10.48550/arXiv.2304.13878}].

\bibitem{KinoshitaNature2006}
T. Kinoshita, T. Wenger and D. S. Weiss, \emph{A quantum Newton's cradle}, Nature \textbf{440}, 900 (2006),
doi:\href{https://doi.org/10.1038/nature04693}{10.1038/nature04693}.

\bibitem{RigolPRL2007}
M. Rigol, V. Dunjko, V. Yurovsky and M. Olshanii, \emph{Relaxation in a completely integrable many-body
quantum system: An ab initio study of the dynamics of the highly excited states of 1D lattice hard-core
bosons}, Phys. Rev. Lett. \textbf{98}, 050405 (2007),
doi:\href{https://doi.org/10.1103/PhysRevLett.98.050405}{10.1103/PhysRevLett.98.050405} [preprint
doi:\href{https://doi.org/10.48550/arXiv.cond-mat/0604476}{10.48550/arXiv.cond-mat/0604476}].

\bibitem{FosterPRL2010}
M. S. Foster, E. A. Yuzbashyan and B. L. Altshuler, \emph{Quantum quench in one dimension: Coherent
inhomogeneity amplification and ``supersolitons''}, Phys. Rev. Lett. \textbf{105}, 135701 (2010),
doi:\href{https://doi.org/10.1103/PhysRevLett.105.135701}{10.1103/PhysRevLett.105.135701} [preprint
doi:\href{https://doi.org/10.48550/arXiv.1006.0012}{10.48550/arXiv.1006.0012}].

\bibitem{PolkovnikovRMP2011}
A. Polkovnikov, K. Sengupta, A. Silva and M. Vengalattore, \emph{Colloquium: Nonequilibrium dynamics of closed
interacting quantum systems}, Rev. Mod. Phys. \textbf{83}, 863 (2011),
doi:\href{https://doi.org/10.1103/RevModPhys.83.863}{10.1103/RevModPhys.83.863}.

\bibitem{FosterPRB2011}
M. S. Foster, T. C. Berkelbach, D. R. Reichman and E. A. Yuzbashyan, \emph{Quantum quench spectroscopy of a
Luttinger liquid: Ultrarelativistic density wave dynamics due to fractionalization in an $XXZ$ chain}, Phys.
Rev. B \textbf{84}, 085146 (2011),
doi:\href{https://doi.org/10.1103/PhysRevB.84.085146}{10.1103/PhysRevB.84.085146} [preprint
doi:\href{https://doi.org/10.48550/arXiv.1105.1786}{10.48550/arXiv.1105.1786}].

\bibitem{MatsunagaShimanoPRL2012}
R. Matsunaga and R. Shimano, \emph{Nonequilibrium BCS state dynamics induced by intense terahertz pulses in a
superconducting NbN film}, Phys. Rev. Lett. \textbf{109}, 187002 (2012),
doi:\href{https://doi.org/10.1103/PhysRevLett.109.187002}{10.1103/PhysRevLett.109.187002}.

\bibitem{BeckPRL2013}
M. Beck, I. Rousseau, M. Klammer, P. Leiderer, M. Mittendorff, S. Winnerl, M. Helm, G. N. Gol'tsman and J.
Demsar, \emph{Transient increase of the energy gap of superconducting NbN thin films excited by resonant
narrow-band terahertz pulses}, Phys. Rev. Lett. \textbf{110}, 267003 (2013),
doi:\href{https://doi.org/10.1103/PhysRevLett.110.267003}{10.1103/PhysRevLett.110.267003}.

\bibitem{CauxEsslerPRL2013}
J.-S. Caux and F. H. L. Essler, \emph{Time evolution of local observables after quenching to an integrable
model}, Phys. Rev. Lett. \textbf{110}, 257203 (2013),
doi:\href{https://doi.org/10.1103/PhysRevLett.110.257203}{10.1103/PhysRevLett.110.257203} [preprint
doi:\href{https://doi.org/10.48550/arXiv.1301.3806}{10.48550/arXiv.1301.3806}].

\bibitem{MatsunagaScience2014}
R. Matsunaga et al., \emph{Light-induced collective pseudospin precession resonating with Higgs mode in a
superconductor}, Science \textbf{345}, 1145 (2014),
doi:\href{https://doi.org/10.1126/science.1254697}{10.1126/science.1254697}.

\bibitem{FosterPRL2014}
M. S. Foster, V. Gurarie, M. Dzero and E. A. Yuzbashyan, \emph{Quench-induced Floquet topological $p$-wave
superfluids}, Phys. Rev. Lett. \textbf{113}, 076403 (2014),
doi:\href{https://doi.org/10.1103/PhysRevLett.113.076403}{10.1103/PhysRevLett.113.076403} [preprint
doi:\href{https://doi.org/10.48550/arXiv.1307.2256}{10.48550/arXiv.1307.2256}].

\bibitem{WoutersPRL2014}
B. Wouters, J. De Nardis, M. Brockmann, D. Fioretto, M. Rigol and J.-S. Caux, \emph{Quenching the anisotropic
Heisenberg chain: Exact solution and generalized Gibbs ensemble predictions}, Phys. Rev. Lett. \textbf{113},
117202 (2014), doi:\href{https://doi.org/10.1103/PhysRevLett.113.117202}{10.1103/PhysRevLett.113.117202}
[preprint doi:\href{https://doi.org/10.48550/arXiv.1405.0172}{10.48550/arXiv.1405.0172}].

\bibitem{YuzbashyanPRA2015}
E. A. Yuzbashyan, M. Dzero, V. Gurarie and M. S. Foster, \emph{Quantum quench phase diagrams of an $s$-wave
BCS-BEC condensate}, Phys. Rev. A \textbf{91}, 033628 (2015),
doi:\href{https://doi.org/10.1103/PhysRevA.91.033628}{10.1103/PhysRevA.91.033628} [preprint
doi:\href{https://doi.org/10.48550/arXiv.1412.7165}{10.48550/arXiv.1412.7165}].

\bibitem{LangenScience2015}
T. Langen et al., \emph{Experimental observation of a generalized Gibbs ensemble}, Science \textbf{348}, 207
(2015), doi:\href{https://doi.org/10.1126/science.1257026}{10.1126/science.1257026} [preprint
doi:\href{https://doi.org/10.48550/arXiv.1411.7185}{10.48550/arXiv.1411.7185}].

\bibitem{IlievskiPRL2015}
E. Ilievski, J. De Nardis, B. Wouters, J.-S. Caux, F. H. L. Essler and T. Prosen, \emph{Complete generalized
Gibbs ensembles in an interacting theory}, Phys. Rev. Lett. \textbf{115}, 157201 (2015),
doi:\href{https://doi.org/10.1103/PhysRevLett.115.157201}{10.1103/PhysRevLett.115.157201} [preprint
doi:\href{https://doi.org/10.48550/arXiv.1507.02993}{10.48550/arXiv.1507.02993}].

\bibitem{EsslerFagotti2016}
F. H. L. Essler and M. Fagotti, \emph{Quench dynamics and relaxation in isolated integrable quantum spin
chains}, J. Stat. Mech.: Theory Exp. 064002 (2016),
doi:\href{https://doi.org/10.1088/1742-5468/2016/06/064002}{10.1088/1742-5468/2016/06/064002} [preprint
doi:\href{https://doi.org/10.48550/arXiv.1603.06452}{10.48550/arXiv.1603.06452}].

\bibitem{VidmarRigol2016}
L. Vidmar and M. Rigol, \emph{Generalized Gibbs ensemble in integrable lattice models}, J. Stat. Mech.: Theory
Exp. 064007 (2016),
doi:\href{https://doi.org/10.1088/1742-5468/2016/06/064007}{10.1088/1742-5468/2016/06/064007} [preprint
doi:\href{https://doi.org/10.48550/arXiv.1604.03990}{10.48550/arXiv.1604.03990}].

\bibitem{CastroAlvaredoPRX2016}
O. A. Castro-Alvaredo, B. Doyon and T. Yoshimura, \emph{Emergent hydrodynamics in integrable quantum systems
out of equilibrium}, Phys. Rev. X \textbf{6}, 041065 (2016),
doi:\href{https://doi.org/10.1103/PhysRevX.6.041065}{10.1103/PhysRevX.6.041065} [preprint
doi:\href{https://doi.org/10.48550/arXiv.1605.07331}{10.48550/arXiv.1605.07331}].

\bibitem{YuzbashyanAOP2016}
E. A. Yuzbashyan, \emph{Generalized microcanonical and Gibbs ensembles in classical and quantum integrable
dynamics}, Ann. Phys. \textbf{367}, 288 (2016),
doi:\href{https://doi.org/10.1016/j.aop.2016.02.002}{10.1016/j.aop.2016.02.002} [preprint
doi:\href{https://doi.org/10.48550/arXiv.1509.06351}{10.48550/arXiv.1509.06351}].

\bibitem{MatsunagaPRB2017}
R. Matsunaga, N. Tsuji, K. Makise, H. Terai, H. Aoki and R. Shimano, \emph{Polarization-resolved terahertz
third-harmonic generation in a single-crystal superconductor NbN: Dominance of the Higgs mode beyond the BCS
approximation}, Phys. Rev. B \textbf{96}, 020505 (2017),
doi:\href{https://doi.org/10.1103/PhysRevB.96.020505}{10.1103/PhysRevB.96.020505} [preprint
doi:\href{https://doi.org/10.48550/arXiv.1703.02815}{10.48550/arXiv.1703.02815}].

\bibitem{ScaramazzaPRB2019}
J. A. Scaramazza, P. Smacchia and E. A. Yuzbashyan, \emph{Consequences of integrability breaking in quench
dynamics of pairing Hamiltonians}, Phys. Rev. B \textbf{99}, 054520 (2019),
doi:\href{https://doi.org/10.1103/PhysRevB.99.054520}{10.1103/PhysRevB.99.054520} [preprint
doi:\href{https://doi.org/10.48550/arXiv.1812.04410}{10.48550/arXiv.1812.04410}].

\bibitem{ShankarPRXQ2022}
A. Shankar, E. A. Yuzbashyan, V. Gurarie, P. Zoller, J. J. Bollinger and A. M. Rey, \emph{Simulating dynamical
phases of chiral $p+ip$ superconductors with a trapped-ion magnet}, PRX Quantum \textbf{3}, 040324 (2022),
doi:\href{https://doi.org/10.1103/PRXQuantum.3.040324}{10.1103/PRXQuantum.3.040324}.

\bibitem{MorvanNature2022}
A. Morvan et al., \emph{Formation of robust bound states of interacting microwave photons}, Nature
\textbf{612}, 240 (2022), doi:\href{https://doi.org/10.1038/s41586-022-05348-y}{10.1038/s41586-022-05348-y}
[preprint doi:\href{https://doi.org/10.48550/arXiv.2206.05254}{10.48550/arXiv.2206.05254}].

\bibitem{YoungNature2024}
D. J. Young et al., \emph{Observing dynamical phases of BCS superconductors in a cavity QED simulator}, Nature
\textbf{625}, 679 (2024), doi:\href{https://doi.org/10.1038/s41586-023-06911-x}{10.1038/s41586-023-06911-x}
[preprint doi:\href{https://doi.org/10.48550/arXiv.2306.00066}{10.48550/arXiv.2306.00066}].

\bibitem{KimPolkovnikovPRB2024}
H. Kim and A. Polkovnikov, \emph{Integrability as an attractor of adiabatic flows}, Phys. Rev. B \textbf{109},
195162 (2024), doi:\href{https://doi.org/10.1103/PhysRevB.109.195162}{10.1103/PhysRevB.109.195162} [preprint
doi:\href{https://doi.org/10.48550/arXiv.2308.09745}{10.48550/arXiv.2308.09745}].

\bibitem{Landau}
L. D. Landau, \emph{A theory of energy transfer. II}, in \emph{Collected papers of L.D. Landau}, Elsevier,
Amsterdam, Netherlands, ISBN 9780080105864 (1965),
doi:\href{https://doi.org/10.1016/B978-0-08-010586-4.50014-6}{10.1016/B978-0-08-010586-4.50014-6}.

\bibitem{zener}
C. Zener, \emph{Non-adiabatic crossing of energy levels}, Proc. R. Soc. A \textbf{137}, 696 (1932),
doi:\href{https://doi.org/10.1098/rspa.1932.0165}{10.1098/rspa.1932.0165}.

\bibitem{majorana}
E. Majorana, \emph{Atomi orientati in campo magnetico variabile}, Nuovo Cim. \textbf{9}, 43 (1932),
doi:\href{https://doi.org/10.1007/BF02960953}{10.1007/BF02960953}.

\bibitem{Stueckelberg}
E. C. G. Stueckelberg, \emph{Theory of inelastic collisions between atoms}, Helv. Phys. Acta \textbf{5}, 369
(1932).

\bibitem{FeynmanHibbs}
R. P. Feynman and A. R. Hibbs, \emph{Quantum mechanics and path integrals}, McGraw--Hill, USA, ISBN
9780070206502 (1965).

\bibitem{MukhanovWinitzki}
V. Mukhanov and S. Winitzki, \emph{Introduction to quantum effects in gravity}, Cambridge University Press,
Cambridge, UK, ISBN 9780521868341 (2007),
doi:\href{https://doi.org/10.1017/CBO9780511809149}{10.1017/CBO9780511809149}.

\bibitem{PerelomovPopov}
A. M. Perelomov and V. S. Popov, \emph{Group-theoretical aspects of the variable frequency oscillator
problem}, Theor. Math. Phys. \textbf{1}, 275 (1969),
doi:\href{https://doi.org/10.1007/BF01035742}{10.1007/BF01035742}.

\bibitem{DemkovOsherov}
Y. N. Demkov and V. I. Osherov, \emph{Stationary and nonstationary problems in quantum mechanics that can be
solved by means of contour integration}, Sov. Phys. J. Exp. Theo. Phys. \textbf{26}, 916 (1968).

\bibitem{OstrovskyNakamura}
V. N. Ostrovsky and H. Nakamura, \emph{Exact analytical solution of the N-level Landau-Zener-type bow-tie
model}, J. Phys. A: Math. Gen. \textbf{30}, 6939 (1997),
doi:\href{https://doi.org/10.1088/0305-4470/30/19/028}{10.1088/0305-4470/30/19/028}.

\bibitem{DemkovOstrovsky2000}
Y. N. Demkov and V. N. Ostrovsky, \emph{Multipath interference in a multistate Landau-Zener-type model}, Phys.
Rev. A \textbf{61}, 032705 (2000),
doi:\href{https://doi.org/10.1103/PhysRevA.61.032705}{10.1103/PhysRevA.61.032705}.

\bibitem{DemkovOstrovsky2001}
Y. N. Demkov and V. N. Ostrovsky, \emph{The exact solution of the multistate Landau-Zener type model: The
generalized bow-tie model}, J. Phys. B: At. Mol. Opt. Phys. \textbf{34}, 2419 (2001),
doi:\href{https://doi.org/10.1088/0953-4075/34/12/309}{10.1088/0953-4075/34/12/309}.

\bibitem{ChernyakSinitsynSun2018}
V. Y. Chernyak, N. A. Sinitsyn and C. Sun, \emph{A large class of solvable multistate Landau-Zener models and
quantum integrability}, J. Phys. A: Math. Theor. \textbf{51}, 245201 (2018),
doi:\href{https://doi.org/10.1088/1751-8121/aac3b2}{10.1088/1751-8121/aac3b2} [preprint
doi:\href{https://doi.org/10.48550/arXiv.1707.04963}{10.48550/arXiv.1707.04963}].

\bibitem{ChernyakSinitsynSun}
V. Y. Chernyak, N. A. Sinitsyn and C. Sun, \emph{Multitime Landau-Zener model: Classification of solvable
Hamiltonians*}, J. Phys. A: Math. Theor. \textbf{53}, 185203 (2020),
doi:\href{https://doi.org/10.1088/1751-8121/ab7fdd}{10.1088/1751-8121/ab7fdd} [preprint
doi:\href{https://doi.org/10.48550/arXiv.1911.12219}{10.48550/arXiv.1911.12219}].

\bibitem{Barik2025}
S. Barik, L. Bakker, V. Gritsev and E. Yuzbashyan, \emph{Knizhnik-Zamolodchikov equations and integrable
hyperbolic Landau-Zener models}, SciPost Phys. \textbf{18}, 212 (2025),
doi:\href{https://doi.org/10.21468/SciPostPhys.18.6.212}{10.21468/SciPostPhys.18.6.212} [preprint
doi:\href{https://doi.org/10.48550/arXiv.2409.17053}{10.48550/arXiv.2409.17053}].

\bibitem{Polkovnikov2005}
A. Polkovnikov, \emph{Universal adiabatic dynamics in the vicinity of a quantum critical point}, Phys. Rev. B
\textbf{72}, 161201 (2005), doi:\href{https://doi.org/10.1103/PhysRevB.72.161201}{10.1103/PhysRevB.72.161201}
[preprint doi:\href{https://doi.org/10.48550/arXiv.cond-mat/0312144}{10.48550/arXiv.cond-mat/0312144}].

\bibitem{ZurekDornerZoller2005}
W. H. Zurek, U. Dorner and P. Zoller, \emph{Dynamics of a quantum phase transition}, Phys. Rev. Lett.
\textbf{95}, 105701 (2005),
doi:\href{https://doi.org/10.1103/PhysRevLett.95.105701}{10.1103/PhysRevLett.95.105701} [preprint
doi:\href{https://doi.org/10.48550/arXiv.cond-mat/0503511}{10.48550/arXiv.cond-mat/0503511}].

\bibitem{Dziarmaga2005}
J. Dziarmaga, \emph{Dynamics of a quantum phase transition: Exact solution of the quantum Ising model}, Phys.
Rev. Lett. \textbf{95}, 245701 (2005),
doi:\href{https://doi.org/10.1103/PhysRevLett.95.245701}{10.1103/PhysRevLett.95.245701} [preprint
doi:\href{https://doi.org/10.48550/arXiv.cond-mat/0509490}{10.48550/arXiv.cond-mat/0509490}].

\bibitem{Galitski2011}
V. Galitski, \emph{Quantum-to-classical correspondence and Hubbard-Stratonovich dynamical systems: A
Lie-algebraic approach}, Phys. Rev. A \textbf{84}, 012118 (2011),
doi:\href{https://doi.org/10.1103/PhysRevA.84.012118}{10.1103/PhysRevA.84.012118} [preprint
doi:\href{https://doi.org/10.48550/arXiv.1012.2873}{10.48550/arXiv.1012.2873}].

\bibitem{RingelGritsev2013}
M. Ringel and V. Gritsev, \emph{Dynamical symmetry approach to path integrals of quantum spin systems}, Phys.
Rev. A \textbf{88}, 062105 (2013),
doi:\href{https://doi.org/10.1103/PhysRevA.88.062105}{10.1103/PhysRevA.88.062105} [preprint
doi:\href{https://doi.org/10.48550/arXiv.1307.2287}{10.48550/arXiv.1307.2287}].

\bibitem{DavidsonPolkovnikov2015}
S. M. Davidson and A. Polkovnikov, \emph{$SU(3)$ semiclassical representation of quantum dynamics of
interacting spins}, Phys. Rev. Lett. \textbf{114}, 045701 (2015),
doi:\href{https://doi.org/10.1103/PhysRevLett.114.045701}{10.1103/PhysRevLett.114.045701} [preprint
doi:\href{https://doi.org/10.48550/arXiv.1406.1192}{10.48550/arXiv.1406.1192}].

\bibitem{PatraYuzbashyan2015}
A. Patra and E. A. Yuzbashyan, \emph{Quantum integrability in the multistate Landau-Zener problem}, J. Phys.
A: Math. Theor. \textbf{48}, 245303 (2015),
doi:\href{https://doi.org/10.1088/1751-8113/48/24/245303}{10.1088/1751-8113/48/24/245303} [preprint
doi:\href{https://doi.org/10.48550/arXiv.1412.4926}{10.48550/arXiv.1412.4926}].

\bibitem{GritsevPolkovnikov}
V. Gritsev and A. Polkovnikov, \emph{Integrable Floquet dynamics}, SciPost Phys. \textbf{2}, 021 (2017),
doi:\href{https://doi.org/10.21468/SciPostPhys.2.3.021}{10.21468/SciPostPhys.2.3.021} [preprint
doi:\href{https://doi.org/10.48550/arXiv.1701.05276}{10.48550/arXiv.1701.05276}].

\bibitem{MalikisCheianov2025}
S. Malikis and V. Cheianov, \emph{Exact S-matrices for higher dimensional representations of generalized
Landau-Zener Hamiltonians}, Phys. Rev. A \textbf{113}, 012201 (2026),
doi:\href{https://doi.org/10.1103/sblf-75mc}{10.1103/sblf-75mc} [preprint
doi:\href{https://doi.org/10.48550/arXiv.2505.06048}{10.48550/arXiv.2505.06048}].

\bibitem{Hirota1979}
R. Hirota, \emph{The bäcklund and inverse scattering transform of the K-dV equation with nonuniformities}, J.
Phys. Soc. Jpn. \textbf{46}, 1681 (1979),
doi:\href{https://doi.org/10.1143/JPSJ.46.1681}{10.1143/JPSJ.46.1681}.

\bibitem{Nirmala1986}
N. Nirmala, M. J. Vedan and B. V. Baby, \emph{Auto-Bäcklund transformation, Lax pairs, and Painlevé property
of a variable coefficient Korteweg-de Vries equation. I}, J. Math. Phys. \textbf{27}, 2640 (1986),
doi:\href{https://doi.org/10.1063/1.527282}{10.1063/1.527282}.

\bibitem{Chan1989}
W. L. Chan and L. Kam-Shun, \emph{Nonpropagating solitons of the variable coefficient and nonisospectral
Korteweg-de Vries equation}, J. Math. Phys. \textbf{30}, 2521 (1989),
doi:\href{https://doi.org/10.1063/1.528533}{10.1063/1.528533}.

\bibitem{GangulyDas2015}
A. Ganguly and A. Das, \emph{Schrödinger equation with time-dependent mass function and associated generalized
KdV equation}, Phys. Scr. \textbf{90}, 055204 (2015),
doi:\href{https://doi.org/10.1088/0031-8949/90/5/055204}{10.1088/0031-8949/90/5/055204}.

\bibitem{Hlavaty1986}
L. Hlavatý, \emph{The Painlevé analysis of damped KdV equation}, J. Phys. Soc. Jpn. \textbf{55}, 1405 (1986),
doi:\href{https://doi.org/10.1143/JPSJ.55.1405}{10.1143/JPSJ.55.1405}.

\bibitem{Joshi1987}
N. Joshi, \emph{Painlevé property of general variable-coefficient versions of the Korteweg-de Vries and
non-linear Schrödinger equations}, Phys. Lett. A \textbf{125}, 456 (1987),
doi:\href{https://doi.org/10.1016/0375-9601(87)90184-8}{10.1016/0375-9601(87)90184-8}.

\bibitem{Brugarino1989}
T. Brugarino, \emph{Painlevé property, auto-Bäcklund transformation, Lax pairs, and reduction to the standard
form for the Korteweg-De Vries equation with nonuniformities}, J. Math. Phys. \textbf{30}, 1013 (1989),
doi:\href{https://doi.org/10.1063/1.528368}{10.1063/1.528368}.

\bibitem{PerezGarcia2006}
V. M. Pérez-García, P. J. Torres and V. V. Konotop, \emph{Similarity transformations for nonlinear Schrödinger
equations with time-dependent coefficients}, Phys. D: Non-linear Phenom. \textbf{221}, 31 (2006),
doi:\href{https://doi.org/10.1016/j.physd.2006.07.002}{10.1016/j.physd.2006.07.002} [preprint
doi:\href{https://doi.org/10.48550/arXiv.nlin/0512061}{10.48550/arXiv.nlin/0512061}].

\bibitem{MalikisCheianov2022}
S. Malikis and V. Cheianov, \emph{An ideal rapid-cycle Thouless pump}, SciPost Phys. \textbf{12}, 203 (2022),
doi:\href{https://doi.org/10.21468/SciPostPhys.12.6.203}{10.21468/SciPostPhys.12.6.203} [preprint
doi:\href{https://doi.org/10.48550/arXiv.2104.02751}{10.48550/arXiv.2104.02751}].

\bibitem{Hoare2020}
B. Hoare, N. Levine and A. A. Tseytlin, \emph{Sigma models with local couplings: A new integrability-RG flow
connection}, J. High Energy Phys. \textbf{11}, 020 (2020),
doi:\href{https://doi.org/10.1007/JHEP11(2020)020}{10.1007/JHEP11(2020)020}.

\bibitem{sinitsyn_integrable_2018}
N. A. Sinitsyn, E. A. Yuzbashyan, V. Y. Chernyak, A. Patra and C. Sun, \emph{Integrable time-dependent quantum
Hamiltonians}, Phys. Rev. Lett. \textbf{120}, 190402 (2018),
doi:\href{https://doi.org/10.1103/PhysRevLett.120.190402}{10.1103/PhysRevLett.120.190402} [preprint
doi:\href{https://doi.org/10.48550/arXiv.1711.09945}{10.48550/arXiv.1711.09945}].

\bibitem{Yuzbashyan2018}
E. A. Yuzbashyan, \emph{Integrable time-dependent Hamiltonians, solvable Landau-Zener models and Gaudin
magnets}, Ann. Phys. \textbf{392}, 323 (2018),
doi:\href{https://doi.org/10.1016/j.aop.2018.01.017}{10.1016/j.aop.2018.01.017} [preprint
doi:\href{https://doi.org/10.48550/arXiv.1802.01571}{10.48550/arXiv.1802.01571}].

\bibitem{Zabalo2022}
A. Zabalo, A.-K. Wu, J. H. Pixley and E. A. Yuzbashyan, \emph{Nonlocality as the source of purely quantum
dynamics of BCS superconductors}, Phys. Rev. B \textbf{106}, 104513 (2022),
doi:\href{https://doi.org/10.1103/PhysRevB.106.104513}{10.1103/PhysRevB.106.104513} [preprint
doi:\href{https://doi.org/10.48550/arXiv.2208.07320}{10.48550/arXiv.2208.07320}].

\bibitem{SuzukiMallaSinitsyn2025arxiv}
F. Suzuki, R. K. Malla and N. A. Sinitsyn, \emph{Competing bosonic reactions: Insight from exactly solvable
time-dependent models}, Phys. Rev. A \textbf{112}, 013307 (2025),
doi:\href{https://doi.org/10.1103/68zs-gv5y}{10.1103/68zs-gv5y} [preprint
doi:\href{https://doi.org/10.48550/arXiv.2504.13027}{10.48550/arXiv.2504.13027}].

\bibitem{GaudinBook}
M. Gaudin, \emph{The Bethe wavefunction}, Cambridge University Press, Cambridge, UK, ISBN 9781107045859
(2014), doi:\href{https://doi.org/10.1017/CBO9781107053885}{10.1017/CBO9781107053885}.

\bibitem{Sklyanin1989}
E. K. Sklyanin, \emph{Separation of variables in the Gaudin model}, J. Sov. Math. \textbf{47}, 2473 (1989),
doi:\href{https://doi.org/10.1007/BF01840429}{10.1007/BF01840429}.

\bibitem{Ortiz2005}
G. Ortiz, R. Somma, J. Dukelsky and S. Rombouts, \emph{Exactly-solvable models derived from a generalized
Gaudin algebra}, Nucl. Phys. B \textbf{707}, 421 (2005),
doi:\href{https://doi.org/10.1016/j.nuclphysb.2004.11.008}{10.1016/j.nuclphysb.2004.11.008} [preprint
doi:\href{https://doi.org/10.48550/arXiv.cond-mat/0407429}{10.48550/arXiv.cond-mat/0407429}].

\bibitem{Skrypnyk2007b}
T. Skrypnyk, \emph{Generalized Gaudin spin chains, nonskew symmetric r-matrices, and reflection equation
algebras}, J. Math. Phys. \textbf{48}, 113521 (2007),
doi:\href{https://doi.org/10.1063/1.2816256}{10.1063/1.2816256}.

\bibitem{Skrypnyk2007}
T. Skrypnyk, \emph{Generalized Gaudin systems in a magnetic field and non-skew-symmetric r-matrices}, J. Phys.
A: Math. Theor. \textbf{40}, 13337 (2007),
doi:\href{https://doi.org/10.1088/1751-8113/40/44/014}{10.1088/1751-8113/40/44/014}.

\bibitem{FiorettoCauxGritsev2014}
D. Fioretto, J.-S. Caux and V. Gritsev, \emph{Exact out-of-equilibrium central spin dynamics from
integrability}, New J. Phys. \textbf{16}, 043024 (2014),
doi:\href{https://doi.org/10.1088/1367-2630/16/4/043024}{10.1088/1367-2630/16/4/043024} [preprint
doi:\href{https://doi.org/10.48550/arXiv.1211.5905}{10.48550/arXiv.1211.5905}].

\bibitem{Barik2025RG}
S. Barik, L. Bakker, V. Gritsev, J. Min\'{a}\v{r} and E. A. Yuzbashyan, \emph{Higher spin Richardson-Gaudin
model with time-dependent coupling: Exact dynamics}, Phys. Rev. B \textbf{113}, 195147 (2026),
doi:\href{https://doi.org/10.1103/c8vm-hz26}{10.1103/c8vm-hz26} [preprint
doi:\href{https://doi.org/10.48550/arXiv.2507.10856}{10.48550/arXiv.2507.10856}].

\bibitem{LiChernyakSinitsyn2018}
F. Li, V. Y. Chernyak and N. A. Sinitsyn, \emph{Quantum annealing and thermalization: Insights from
integrability}, Phys. Rev. Lett. \textbf{121}, 190601 (2018),
doi:\href{https://doi.org/10.1103/PhysRevLett.121.190601}{10.1103/PhysRevLett.121.190601} [preprint
doi:\href{https://doi.org/10.48550/arXiv.1804.00371}{10.48550/arXiv.1804.00371}].

\bibitem{Malla2022}
R. K. Malla, V. Y. Chernyak, C. Sun and N. A. Sinitsyn, \emph{Coherent reaction between molecular and atomic
Bose-Einstein condensates: Integrable model}, Phys. Rev. Lett. \textbf{129}, 033201 (2022),
doi:\href{https://doi.org/10.1103/PhysRevLett.129.033201}{10.1103/PhysRevLett.129.033201} [preprint
doi:\href{https://doi.org/10.48550/arXiv.2112.12302}{10.48550/arXiv.2112.12302}].

\bibitem{SinitsynLi2016}
N. A. Sinitsyn and F. Li, \emph{Solvable multistate model of Landau-Zener transitions in cavity QED}, Phys.
Rev. A \textbf{93}, 063859 (2016),
doi:\href{https://doi.org/10.1103/PhysRevA.93.063859}{10.1103/PhysRevA.93.063859} [preprint
doi:\href{https://doi.org/10.48550/arXiv.1602.03136}{10.48550/arXiv.1602.03136}].

\bibitem{SunSinitsyn2016}
C. Sun and N. A. Sinitsyn, \emph{Landau-Zener extension of the Tavis-Cummings model: Structure of the
solution}, Phys. Rev. A \textbf{94}, 033808 (2016),
doi:\href{https://doi.org/10.1103/PhysRevA.94.033808}{10.1103/PhysRevA.94.033808} [preprint
doi:\href{https://doi.org/10.48550/arXiv.1606.08430}{10.48550/arXiv.1606.08430}].

\bibitem{Sun2019}
C. Sun, V. Y. Chernyak, A. Piryatinski and N. A. Sinitsyn, \emph{Cooperative light emission in the presence of
strong inhomogeneous broadening}, Phys. Rev. Lett. \textbf{123}, 123605 (2019),
doi:\href{https://doi.org/10.1103/PhysRevLett.123.123605}{10.1103/PhysRevLett.123.123605} [preprint
doi:\href{https://doi.org/10.48550/arXiv.1906.11412}{10.48550/arXiv.1906.11412}].

\bibitem{Torrielli2016}
A. Torrielli, \emph{Classical integrability}, J. Phys. A: Math. Theor. \textbf{49}, 323001 (2016),
doi:\href{https://doi.org/10.1088/1751-8113/49/32/323001}{10.1088/1751-8113/49/32/323001}.

\bibitem{Loebbert2016}
F. Loebbert, \emph{Lectures on Yangian symmetry}, J. Phys. A: Math. Theor. \textbf{49}, 323002 (2016),
doi:\href{https://doi.org/10.1088/1751-8113/49/32/323002}{10.1088/1751-8113/49/32/323002} [preprint
doi:\href{https://doi.org/10.48550/arXiv.1606.02947}{10.48550/arXiv.1606.02947}].

\bibitem{Retore2022}
A. L. Retore, \emph{Introduction to classical and quantum integrability}, J. Phys. A: Math. Theor.
\textbf{55}, 173001 (2022), doi:\href{https://doi.org/10.1088/1751-8121/ac5a8e}{10.1088/1751-8121/ac5a8e}
[preprint doi:\href{https://doi.org/10.48550/arXiv.2109.14280}{10.48550/arXiv.2109.14280}].

\bibitem{McGuire1964}
J. B. McGuire, \emph{Study of exactly soluble one-dimensional N-body problems}, J. Math. Phys. \textbf{5}, 622
(1964), doi:\href{https://doi.org/10.1063/1.1704156}{10.1063/1.1704156}.

\bibitem{Yang1967}
C. N. Yang, \emph{Some exact results for the many-body problem in one dimension with repulsive delta-function
interaction}, Phys. Rev. Lett. \textbf{19}, 1312 (1967),
doi:\href{https://doi.org/10.1103/PhysRevLett.19.1312}{10.1103/PhysRevLett.19.1312}.

\bibitem{Zamolodchikov1979}
A. B. Zamolodchikov and A. B. Zamolodchikov, \emph{Factorized S-matrices in two dimensions as the exact
solutions of certain relativistic quantum field theory models}, Ann. Phys. \textbf{120}, 253 (1979),
doi:\href{https://doi.org/10.1016/0003-4916(79)90391-9}{10.1016/0003-4916(79)90391-9}.

\bibitem{Baxter1982}
R. J. Baxter, \emph{Exactly solved models in statistical mechanics}, Academic Press, London, UK, ISBN
9780120831807 (1982).

\bibitem{Jimbo1989}
M. Jimbo, \emph{Introduction to the Yang-Baxter equation}, Int. J. Mod. Phys. A \textbf{04}, 3759 (1989),
doi:\href{https://doi.org/10.1142/S0217751X89001503}{10.1142/S0217751X89001503}.

\bibitem{Parmesh1}
P. R. Pasnoori, \emph{Integrability of the Kondo model with time-dependent interaction strength}, Phys. Rev. B
\textbf{112}, L060409 (2025), doi:\href{https://doi.org/10.1103/78xb-5lmw}{10.1103/78xb-5lmw}.

\bibitem{Anderson}
P. W. Anderson, G. Yuval and D. R. Hamann, \emph{Exact results in the Kondo problem. II. Scaling theory,
qualitatively correct solution, and some new results on one-dimensional classical statistical models}, Phys.
Rev. B \textbf{1}, 4464 (1970), doi:\href{https://doi.org/10.1103/PhysRevB.1.4464}{10.1103/PhysRevB.1.4464}.

\bibitem{Wilson}
K. G. Wilson, \emph{The renormalization group: Critical phenomena and the Kondo problem}, Rev. Mod. Phys.
\textbf{47}, 773 (1975), doi:\href{https://doi.org/10.1103/RevModPhys.47.773}{10.1103/RevModPhys.47.773}.

\bibitem{Nozier}
Ph. Nozières and A. Blandin, \emph{Kondo effect in real metals}, J. Phys. \textbf{41}, 193 (1980),
doi:\href{https://doi.org/10.1051/jphys:01980004103019300}{10.1051/jphys:01980004103019300}.

\bibitem{Andrei80}
N. Andrei, \emph{Diagonalization of the Kondo Hamiltonian}, Phys. Rev. Lett. \textbf{45}, 379 (1980),
doi:\href{https://doi.org/10.1103/PhysRevLett.45.379}{10.1103/PhysRevLett.45.379}.

\bibitem{Wiegmann_1981}
P. B. Wiegmann, \emph{Exact solution of the s-d exchange model (Kondo problem)}, J. Phys. C: Solid State Phys.
\textbf{14}, 1463 (1981), doi:\href{https://doi.org/10.1088/0022-3719/14/10/014}{10.1088/0022-3719/14/10/014}.

\bibitem{Cherednik1989}
I. V. Cherednik, \emph{Generalized braid groups and local r-matrix systems}, Dokl. Akad. Nauk SSSR
\textbf{307}, 49 (1989).

\bibitem{Etingof1994}
P. I. Etingof, \emph{Representations of affine Lie algebras, ellipticr-matrix systems, and special functions},
Commun. Math. Phys. \textbf{159}, 471 (1994),
doi:\href{https://doi.org/10.1007/BF02099981}{10.1007/BF02099981} [preprint
doi:\href{https://doi.org/10.48550/arXiv.hep-th/9303018}{10.48550/arXiv.hep-th/9303018}].

\bibitem{BelavinDrinfeld1982}
A. A. Belavin and V. G. Drinfel'd, \emph{Solutions of the classical Yang - Baxter equation for simple Lie
algebras}, Funct. Anal. Appl. \textbf{16}, 159 (1982),
doi:\href{https://doi.org/10.1007/BF01081585}{10.1007/BF01081585}.

\bibitem{BelavinDrinfeld1984}
A. A. Belavin and V. G. Drinfeld, \emph{Triangle equations for simple Lie algebras}, Harwood Academic
Publishers, Amsterdam, Netherlands, ISBN 9789057022692 (1984).

\bibitem{Stolin1991}
A. Stolin, \emph{On rational solutions of Yang-Baxter equation for $\mathrm{sl}(n)$.}, Math. Scand.
\textbf{69}, 57 (1991), doi:\href{https://doi.org/10.7146/math.scand.a-12369}{10.7146/math.scand.a-12369}.

\bibitem{Hikami1995}
K. Hikami, \emph{Gaudin magnet with boundary and generalized Knizhnik-Zamolodchikov equation}, J. Phys. A:
Math. Gen. \textbf{28}, 4997 (1995),
doi:\href{https://doi.org/10.1088/0305-4470/28/17/028}{10.1088/0305-4470/28/17/028}.

\bibitem{BabujianKitaev1998}
H. M. Babujian and A. V. Kitaev, \emph{Generalized Knizhnik-Zamolodchikov equations and isomonodromy
quantization of the equations integrable via the inverse scattering transform: Maxwell-Bloch system with
pumping}, J. Math. Phys. \textbf{39}, 2499 (1998),
doi:\href{https://doi.org/10.1063/1.532404}{10.1063/1.532404}.

\bibitem{KurakLimaSantos2004}
V. Kurak and A. Lima-Santos, \emph{The $A_2^{(2)}$ Gaudin model and its associated Knizhnik-Zamolodchikov
equation}, J. Phys. A: Math. Gen. \textbf{38}, 333 (2004),
doi:\href{https://doi.org/10.1088/0305-4470/38/2/004}{10.1088/0305-4470/38/2/004} [preprint
doi:\href{https://doi.org/10.48550/arXiv.nlin/0407041}{10.48550/arXiv.nlin/0407041}].

\bibitem{LimaSantosUtiel2006}
A. Lima-santos and W. Utiel, \emph{Gaudin magnet with impurity and its generalized Knizhnik-Zamolodchikov
equation}, Int. J. Mod. Phys. B \textbf{20}, 2175 (2006),
doi:\href{https://doi.org/10.1142/S0217979206034595}{10.1142/S0217979206034595} [preprint
doi:\href{https://doi.org/10.48550/arXiv.nlin/0504060}{10.48550/arXiv.nlin/0504060}].

\bibitem{SedrakyanGalitski2010}
T. A. Sedrakyan and V. Galitski, \emph{Boundary Wess-Zumino-Novikov-Witten model from the pairing
Hamiltonian}, Phys. Rev. B \textbf{82}, 214502 (2010),
doi:\href{https://doi.org/10.1103/PhysRevB.82.214502}{10.1103/PhysRevB.82.214502} [preprint
doi:\href{https://doi.org/10.48550/arXiv.1005.0544}{10.48550/arXiv.1005.0544}].

\bibitem{Skrypnyk2010}
T. Skrypnyk, \emph{Generalized Knizhnik-Zamolodchikov equations, off-shell Bethe Ansatz and non-skew-symmetric
classical r-matrices}, Nucl. Phys. B \textbf{824}, 436 (2010),
doi:\href{https://doi.org/10.1016/j.nuclphysb.2009.08.005}{10.1016/j.nuclphysb.2009.08.005}.

\bibitem{KnizhnikZamolodchikov1984}
V. G. Knizhnik and A. B. Zamolodchikov, \emph{Current algebra and Wess-Zumino model in two dimensions}, Nucl.
Phys. B \textbf{247}, 83 (1984),
doi:\href{https://doi.org/10.1016/0550-3213(84)90374-2}{10.1016/0550-3213(84)90374-2}.

\bibitem{Frenkel}
I. B. Frenkel and N. Y. Reshetikhin, \emph{Quantum affine algebras and holonomic difference equations},
Commun. Math. Phys. \textbf{146}, 1 (1992), doi:\href{https://doi.org/10.1007/BF02099206}{10.1007/BF02099206}.

\bibitem{rishetikhin1}
N. Reshetikhin, \emph{Jackson-type integrals, Bethe vectors, and solutions to a difference analog of the
Knizhnik-Zamolodchikov system}, Lett. Math. Phys. \textbf{26}, 153 (1992),
doi:\href{https://doi.org/10.1007/BF00420749}{10.1007/BF00420749}.

\bibitem{smirnovqKZ}
F. A. SMIRNOV, \emph{Dynamical symmetries of massive integrable models 1: Form factor bootstrap equations as a
special case of deformed Knizhnik-Zamolodchikov equations}, Int. J. Mod. Phys. A \textbf{07}, 813 (1992),
doi:\href{https://doi.org/10.1142/S0217751X92004063}{10.1142/S0217751X92004063}.

\bibitem{smirnovqkz2}
F. A. Smirnov, \emph{Completely integrable models of quantum field theory}, in \emph{Form factors in
completely integrable models of quantum field theory}, World Scientific, Singapore, ISBN 9789812798312 (1992),
doi:\href{https://doi.org/10.1142/9789812798312_0001}{10.1142/9789812798312\_0001}.

\bibitem{JimboqKZ}
M. Jimbo and T. Miwa, \emph{Algebraic analysis of solvable lattice models}, American Mathematical Society,
Providence, USA, ISBN 9780821803202 (1994).

\bibitem{Babujian_1997}
H. Babujian, M. Karowski and A. Zapletal, \emph{Matrix difference equations and a nested Bethe Ansatz}, J.
Phys. A: Math. Gen. \textbf{30}, 6425 (1997),
doi:\href{https://doi.org/10.1088/0305-4470/30/18/019}{10.1088/0305-4470/30/18/019} [preprint
doi:\href{https://doi.org/10.48550/arXiv.hep-th/9611006}{10.48550/arXiv.hep-th/9611006}].

\bibitem{Etingof1998}
P. Etingof and O. Schiffmann, \emph{Lectures on the dynamical Yang-Baxter equations}, in \emph{Quantum groups
and lie theory}, Cambridge University Press, Cambridge, UK, ISBN 9780521010405 (2002),
doi:\href{https://doi.org/10.1017/CBO9780511542848.007}{10.1017/CBO9780511542848.007}.

\bibitem{Nakayashiki1999}
A. Nakayashiki, S. Pakuliak and V. Tarasov, \emph{On solutions of the KZ and qKZ equations at level zero},
Ann. Inst. H. Poincaré Phys. Théor. \textbf{71}, 459 (1999) [preprint
doi:\href{https://doi.org/10.48550/arXiv.q-alg/9712002}{10.48550/arXiv.q-alg/9712002}].

\bibitem{Semenov1985}
M. A. Semenov-Tyan-Shanskii, \emph{Classical r-matrices and quantization}, J. Sov. Math. \textbf{31}, 3411
(1985), doi:\href{https://doi.org/10.1007/BF02107242}{10.1007/BF02107242}.

\bibitem{Andrei1995}
N. Andrei, \emph{Integrable models in condensed matter physics}, in \emph{Low-dimensional quantum field
theories for condensed matter physicists}, World Scientific, Singapore, ISBN 9789814447027 (1995),
doi:\href{https://doi.org/10.1142/9789814447027_0008}{10.1142/9789814447027\_0008} [preprint
doi:\href{https://doi.org/10.48550/arXiv.cond-mat/9408101}{10.48550/arXiv.cond-mat/9408101}].

\bibitem{Grobis2007}
M. Grobis, I. G. Rau, R. M. Potok, H. Shtrikman and D. Goldhaber-Gordon, \emph{Universal scaling in
nonequilibrium transport through a single channel kondo dot}, Phys. Rev. Lett. \textbf{100}, 246601 (2008),
doi:\href{https://doi.org/10.1103/PhysRevLett.100.246601}{10.1103/PhysRevLett.100.246601} [preprint
doi:\href{https://doi.org/10.48550/arXiv.0710.3211}{10.48550/arXiv.0710.3211}].

\bibitem{Bauer2013}
J. Bauer, C. Salomon and E. Demler, \emph{Realizing a kondo-correlated state with ultracold atoms}, Phys. Rev.
Lett. \textbf{111}, 215304 (2013),
doi:\href{https://doi.org/10.1103/PhysRevLett.111.215304}{10.1103/PhysRevLett.111.215304} [preprint
doi:\href{https://doi.org/10.48550/arXiv.1308.0603}{10.48550/arXiv.1308.0603}].

\bibitem{Amaricci2025}
A. Amaricci, A. Richaud, M. Capone, N. D. Oppong and F. Scazza, \emph{Engineering the Kondo impurity problem
with alkaline-earth atom arrays}, Phys. Rev. A \textbf{112}, 043301 (2025),
doi:\href{https://doi.org/10.1103/m7l3-y2f8}{10.1103/m7l3-y2f8} [preprint
doi:\href{https://doi.org/10.48550/arXiv.2505.14630}{10.48550/arXiv.2505.14630}].

\bibitem{EsslerKonik2005}
F. H. L. Essler and R. M. Konik, \emph{Application of massive integrable quantum field theories to problems in
condensed matter physics}, in \emph{From fields to strings: Circumnavigating theoretical physics}, World
scientific, Singapore, ISBN 9789812389558 (2005),
doi:\href{https://doi.org/10.1142/9789812775344_0020}{10.1142/9789812775344\_0020} [preprint
doi:\href{https://doi.org/10.48550/arXiv.cond-mat/0412421}{10.48550/arXiv.cond-mat/0412421}].

\bibitem{ThackerRMP1981}
H. B. Thacker, \emph{Exact integrability in quantum field theory and statistical systems}, Rev. Mod. Phys.
\textbf{53}, 253 (1981), doi:\href{https://doi.org/10.1103/RevModPhys.53.253}{10.1103/RevModPhys.53.253}.

\bibitem{Gross1974}
D. J. Gross and A. Neveu, \emph{Dynamical symmetry breaking in asymptotically free field theories}, Phys. Rev.
D \textbf{10}, 3235 (1974), doi:\href{https://doi.org/10.1103/PhysRevD.10.3235}{10.1103/PhysRevD.10.3235}.

\bibitem{Andrei1979}
N. Andrei and J. H. Lowenstein, \emph{Diagonalization of the Chiral-invariant Gross-Neveu Hamiltonian}, Phys.
Rev. Lett. \textbf{43}, 1698 (1979),
doi:\href{https://doi.org/10.1103/PhysRevLett.43.1698}{10.1103/PhysRevLett.43.1698}.

\bibitem{Destri1982}
C. Destri and J. H. Lowenstein, \emph{Analysis of the Bethe-Ansatz equations of the chiral-invariant
Gross-Neveu model}, Nucl. Phys. B \textbf{205}, 369 (1982),
doi:\href{https://doi.org/10.1016/0550-3213(82)90363-7}{10.1016/0550-3213(82)90363-7}.

\bibitem{Thirring1958}
W. E. Thirring, \emph{A soluble relativistic field theory}, Ann. Phys. \textbf{3}, 91 (1958),
doi:\href{https://doi.org/10.1016/0003-4916(58)90015-0}{10.1016/0003-4916(58)90015-0}.

\bibitem{Coleman1975}
S. Coleman, \emph{Quantum sine-Gordon equation as the massive Thirring model}, Phys. Rev. D \textbf{11}, 2088
(1975), doi:\href{https://doi.org/10.1103/PhysRevD.11.2088}{10.1103/PhysRevD.11.2088}.

\bibitem{Bergknoff1979}
H. Bergknoff and H. B. Thacker, \emph{Structure and solution of the massive Thirring model}, Phys. Rev. D
\textbf{19}, 3666 (1979), doi:\href{https://doi.org/10.1103/PhysRevD.19.3666}{10.1103/PhysRevD.19.3666}.

\bibitem{Korepin1979}
V. E. Korepin, \emph{Direct calculation of the S matrix in the massive Thirring model}, Theor. Math. Phys.
\textbf{41}, 953 (1979), doi:\href{https://doi.org/10.1007/BF01028501}{10.1007/BF01028501}.

\bibitem{SklyaninQISM}
E. K. Sklyanin, L. A. Takhtadzhyan and L. D. Faddeev, \emph{Quantum inverse problem method. I}, Theor. Math.
Phys. \textbf{40}, 688 (1979), doi:\href{https://doi.org/10.1007/BF01018718}{10.1007/BF01018718}.

\bibitem{ODBA}
Y. Wang, W.-L. Yang, J. Cao and K. Shi, \emph{Off-Diagonal Bethe Ansatz for exactly solvable models},
Springer, Berlin, Germany, ISBN 9783662467558 (2015),
doi:\href{https://doi.org/10.1007/978-3-662-46756-5}{10.1007/978-3-662-46756-5}.

\end{thebibliography}
\end{document}